%% file: main.tex
\documentclass[preprint,12pt]{elsarticle}

\usepackage{amssymb}
\usepackage{amsmath}
\usepackage[export]{adjustbox}
\usepackage{float}
\usepackage{bm}
\usepackage{booktabs}
\usepackage{comment}
\usepackage{soul}
\usepackage{subfiles} 

\usepackage[dvipsnames]{xcolor}

\usepackage[T1]{fontenc}
\usepackage[utf8]{inputenc}
\usepackage[french,english]{babel}

\definecolor{airforceblue}{rgb}{0.36, 0.54, 0.66}
\definecolor{darkyellow}{rgb}{0.54509804, 0.50196078, 0.}
\newcommand{\rom}[1]{\uppercase\expandafter{\romannumeral #1\relax}}

\newcommand{\HL}[1]{\textcolor{darkyellow}{HL: #1}}

\DeclareMathOperator{\EX}{\mathbb{E}}

\begin{document}

\begin{frontmatter}



\title{The effect of local stacking fault fluctuations on dislocation glide and the interplay with solid solution strengthening in Al-Mg alloys}

\author[clemson]{Hyunsoo Lee} 

\author[Tsinghua]{Xin Liu} 

\author[clemson]{Liam Myhill} 

\author[Shanghai]{Yang Li} 

\author[UL]{David Rodney} 

\author[CNRS,Japan]{Pierre-Antoine Geslin} 

\author[UIUC]{Nikhil Chandra Admal} 

\author[Tsinghua]{Yinan Cui} 

\author[miami]{Giacomo Po} 

\author[clemson,clemson2]{Enrique Martinez} 
\ead{enrique@clemson.edu}

\affiliation[clemson]{
    department={School of Mechanical and Automotive Engineering},
    organization={Clemson University},
    city={Clemson},
    postcode={29634}, 
    state={South Carolina},
    country={USA}
}

\affiliation[clemson2]{
    department={Department of Material Science and Engineering},
    organization={Clemson University},
    city={Clemson},
    postcode={29634}, 
    state={South Carolina},
    country={USA}
}
\affiliation[Tsinghua]{
    department={Department of Engineering Mechanics}, 
    organization={Tsinghua University},
    city={Beijing},
    postcode={100084},
    country={China}
}

\affiliation[Shanghai]{
    department={School of Mechanics and Engineering Science}, 
    organization={Shanghai University},
    city={Shanghai},
    postcode={200444},
    country={China}
}
            
\affiliation[UL]{
    department={Institut Lumi\`ere Mati\`ere, UMR5306},
    organization={Universit\'e Claude Bernard Lyon 1},
    city={Villeurbanne},
    postcode={69622}, 
    country={France}
}

\affiliation[CNRS]{
    department={MATEIS, UMR5510},
    organization={CNRS, INSA Lyon, Universit\'e Claude Bernard Lyon 1},
    organization={INSA Lyon},
    city={Villeurbanne},
    postcode={69621}, 
    country={France}
}

\affiliation[Japan]{
    department={ELyTMaX IRL3757, CNRS, Univ. Lyon, INSA Lyon, Centrale Lyon, Université Claude Bernard Lyon 1},
    organization={Tohoku University},
    city={Sendai},
    postcode={9808577}, 
    country={Japan}
}

\affiliation[UIUC]{
    department={Department of Mechanical Science and Engineering}, 
    organization={University of Illinois Urbana-Champaign},
    city={Urbana},
    state={IL}, 
    postcode={61801},
    country={USA}
}


\affiliation[miami]{
    department={Department of Mechanical and Aerospace Engineering},
    organization={University of Miami}, 
    city={Coral Gables},
    state={FL}, 
    postcode={33146},
    country={USA}
}

\begin{abstract}
Concentrated alloys are characterized by a random arrangement of solute atoms of different sizes, which generates spatially fluctuating displacement and stress fields with non-trivial correlations. This chemical disorder also induces local fluctuations in stacking-fault (SF) energies, giving rise to fluctuating SF forces acting on the partial dislocations bounding the SF. In this work, we investigate the characteristics of the SF noise, including spatial correlations and the impact of such SF energy fluctuations on dislocation glide. Specifically, the gliding behavior of screw, edge, and mixed dislocations in a model concentrated Al-Mg random alloy with varying Mg content is analyzed. The interplay between SF energy fluctuations and solid-solution noise is also discussed. We show that SF forces are spatially correlated, in close analogy with long-range elastic stresses, and exhibit a robust negative correlation along the slip direction. We identify the physical origin of these correlations and demonstrate that they significantly contribute to the critical resolved shear stress for dislocations to glide. In contrast, correlations in other directions are weaker but can still be non-negligible, depending on the underlying interatomic interactions.
\end{abstract}



\begin{highlights}
\item Local stacking fault energies present spatial correlations that affect the critical resolved shear stress for dislocations to glide
\item A broken-bond model is developed to rationalize the origin of the correlations in the stacking fault energy
\item A mesoscale dislocation dynamics model has been developed to consider both stacking fault energy and solid solution fluctuations and their effect on dislocation motion
\end{highlights}


\end{frontmatter}


\section{Introduction}\label{sec:Intro}

\subfile{sections/introduction}

\section{Methodology}\label{sec:Methodology}

\subfile{sections/methodology}

\section{Results}\label{sec:Results}

\subfile{sections/results}

\section{Discussion}\label{sec:Discussion}

\subfile{sections/discussion}

\section{Conclusion}\label{sec:Conclusion}

\subfile{sections/conclusion}

\section*{Acknowledgments}
H.L. and E.M acknowledge the support from the US Department of
Energy, Office of Science, Office of Fusion Energy Sciences under grant number DE-SC0024515. N.A. would like to acknowledge support from the National Science Foundation (NSF), CAREER grant MOMS-2239734. G.P. acknowledges the support of the US National Science Foundation through the CAREER grant number 2340174, and the support of the US Department of Energy, Office of Fusion Energy Sciences, under Awards Number DE-SC0024675 and DE-SC0024401 with the University of Miami.

Additionally, this material is based on
work supported by the National Science Foundation under Grant Nos.
MRI\# 2024205, MRI\# 1725573, and CRI\# 2010270 for allotment of
compute time on the Clemson University Palmetto Cluster.

\section{Appendix}

\subfile{sections/appendix}

\bibliographystyle{ieeetr}
\bibliography{main}

\end{document}

%% file: sections/introduction.tex


Dislocations are main carriers of plasticity in metallic systems and hence the material deformation pathways depend on dislocation glide and climb properties. Understanding the basic mechanisms for dislocations to move might lead to mechanical properties optimization. Concentrated multicomponent alloys have recently become materials of interest since some compositions show outstanding mechanical performance \cite{george2020}. However, these high concentration mixtures pose challenges for dislocation theory to predict critical resolved shear stresses (CRSS) that relate to plastic deformation and strain hardening. The energy landscape for dislocations to move in such complex alloys is significantly rough, owing to the lattice mismatch and differing chemical affinity between the different elements. Different approaches have been developed to understand the effect of composition on dislocation dynamics, starting from Mott and Nabarro~\cite{mott1940attempt}. The Mott-Nabarro model considers the distortion of the matrix due to dilute solid solution (SS) atoms and computes the average magnitude of the stress within the matrix. Later, Friedel and Fleischer improved the model by accounting for the size and modulus mismatch effects of solute atoms~\cite{fleisgher1961solution}. They proposed a statistical model based on the solute concentration $c$, dislocation line tension $\Gamma$, obstacle strength $f_0$, and the average distance between obstacles in the glide plane. The model suggests that the athermal CRSS follows the scaling law $\tau_c \propto \frac{f_0^{3/2} \sqrt{c}} {\sqrt{\Gamma}}$~\cite{varvenne2016theory, fleisgher1961solution, friedel2013dislocations}.
The limitation of the Friedel–Fleischer model is that it cannot properly capture the long-range interactions between dislocations and solutes, as solutes are treated as point-like obstacles. Consequently, it fails to account for concentration fluctuations in near-equiatomic solid solutions, restricting its applicability to the dilute regime~\cite{friedel2013dislocations, fleisgher1961solution}. To overcome the limitations of previous approaches, Labusch proposed the so-called collective pinning theory—a statistical model for dislocation strengthening that accounts for the distribution of interaction forces and characteristic lengths between neighboring obstacles along a dislocation line~\cite{labusch1970statistical}. Labusch made two critical assumptions: (1) solutes outside the glide plane do not interact with dislocations, and (2) the interaction between dislocations and solutes has a cutoff range and a maximum strength $f_0$. The athermal CRSS scaling law proposed by his model gives $\tau_c \propto \frac{c^{2/3} f_0^{4/3}}{\Gamma^{1/3}}$.
This pioneer model is limited because it conflicts with the long-range nature of dislocation–solute interactions. 
Rather, the $\tau_{CRSS}$ scales with the maximum pinning stresses induced by the atomic size mismatch. Decades later, Curtin's group developed a solid solution strengthening model based on the elastic interaction energy between alloy components and dislocations~\cite{barabasi1995fractal,leyson2010quantitative,varvenne2016theory,varvenne2018predicting,larosa2019solid}.

The glide of a dislocation through varying solute atoms within a glide plane has also been modeled as an elastic line evolving in a quenched noise field ~\cite{sevillano1993patterns, d1997internal}. If this noise field is induced by the mismatch of atomic radii imposing local eigenstrains, the noise shall be referred to as solid solution noise. One of the pitfalls of the investigations on the influence of solid solution noise was that they mainly focused on uncorrelated noise~\cite{nattermann1992dynamics, amaral1994universality, olami1995interface, bolech2004universal, kolton2006short}, which tends to neglect systematic strengthening mechanisms due to underlying correlated fields. Previous theoretical works using a random field approach revealed that spatial correlations influence the dislocation glide and transition strength~\cite{park2003dynamics, patinet2013quantitative}.
Recently, Geslin and Rodney \cite{geslin2021microelasticity, geslin2021microelasticity2} developed an analytical theory based on linear elasticity to calculate the stress and displacement fields induced at every point in the material due to the atomic mismatch between different elements. The model relies on the Eshelby inclusion theory and concludes that stresses are auto-correlated. Rida et al. \cite{rida2022influence} showed that considering the effect of such correlations on the dislocation motion is crucial to capture $(\tau_{CRSS})$ for dislocations to glide, which drives the dislocation past local pinning points. Rida also revealed the dependence of $\tau_{CRSS}$ on the atomic composition for an AlMg system, which will be used in this study. Sboui et al. later expanded the solid solution model to consider inclusions featuring non-dilatational components of the eigenstrain as opposed to the original purely dilatation approach \cite{sboui2023elastic}. This effectively expanded the model to consider Elshelby inclusions of various ovoidal shapes in a homogeneous elastic matrix~\cite{sboui2023elastic}. 
Another aspect of complex alloy systems that has typically been overlooked is noise induced by fluctuations in the intrinsic stacking fault (SF) energy $(\gamma_{ISF})$. In metallic alloys with certain crystalline structures, such as face-centered cubic (FCC) or hexagonal close-packed (HCP), full dislocations usually split into partial dislocations \cite{martinez2008atomistically, jo2014theory, kibey2007predicting} based on Frank's rule, leaving an intrinsic SF between them. As a partial dislocation moves, it either generates or removes SF area. The SF energy of that area depends on local composition, and hence, in concentrated systems, it becomes a stochastic variable with a probability distribution. Therefore, the attractive force induced between partial dislocations fluctuates along the dislocation line. Some investigations have been conducted on this subject. Zeng et al. approached the problem by varying the SF energy locally to model the energy landscape of high entropy alloys (HEAs) and demonstrated that their phase field dislocation model agreed well with observations~\cite{zeng2019effects}. Shih et al. showed that the fluctuations in the SF energy cannot be fully explained by previously established solid solution models, claiming that a thermally activated mechanism is at play~\cite{shih2021stacking}. Additionally, Utt et al. demonstrated that the SF energy landscape in HEAs is an important component of the Peierls barrier and a source of jerky dislocation gliding motion~\cite{utt2022origin}. Lastly, Chen et al. made an attempt to model the SF noise by adopting a continuum random field function~\cite{chen2024dislocation}. LaRosa et al. demonstrated that the local SF energy (SFE) can be computed based on bond energies~\cite{larosa2022local}.
A recent work by Vaid et al. adopted the dilatational eigenstrain model~\cite{mura2013micromechanics} to account for the noise of the solid solution in extended dislocation configurations with randomly fluctuating SFE ~\cite{vaid2022pinning}. They revealed that the varying width of the SF under shear stress aligns well with the elastic manifold theory in random fields, a theory originally proposed by Labusch~\cite{labusch1970statistical}. Despite previous efforts to characterize the effect of the SF noise on dislocation glide, the analysis of correlations in the driving forces induced by fluctuations in $\gamma_{ISF}$ has not yet been investigated.

Spatial correlations of solute eigenstrains exert a significant effect on the CRSS~\cite{geslin2021microelasticity, geslin2021microelasticity2}. This work addresses a couple of natural follow-up questions: Does locally fluctuating SFE affect CRSS in random solid solutions? Are these fluctuations correlated? These questions are systematically investigated by exploring the parameters pertinent to dislocation glide under SFE fluctuations. The aim is to establish a general framework to be applicable to a discrete dislocation dynamics model, such that the results can eventually be upscaled to study alloying effects in bulk crystal plasticity. The overview of this work is the following: (1) A general method to extract local SF energies from atomistic simulations is discussed and a framework to construct $\gamma _{ISF}$ spatial correlations for AlMg$_x$ alloy at the mesoscale is proposed. (2) We explore the varying CRSS under different SFE correlations. The locally fluctuating SFE on the glide plane induces fluctuating forces on partial dislocation segments. (3) We further examine dislocation character dependence in the interaction between correlated SFE noise and partial dislocations, and the shape of the dislocations. Simulations demonstrate that distinct noise-dislocation interaction mechanisms yield different CRSS. (4) The origin of the observed negative correlations in $\gamma _{ISF}$ is analytically derived relying on a broken-bond model. (5) The interplay between SFE fluctuations and SS noise is also investigated, assuming superposition holds. Understanding the micromechanical strengthening at play in multicomponent alloys is critical to aid design concentrated engineering alloys for extreme environments. 


%% file: sections/methodology.tex
We use a combined molecular dynamics (MD) and discrete dislocation dynamics (DDD) methodology to conduct our investigation. In DDD, dislocations are discretized in continuum space by nodes, which represent the degrees of freedom, and quadrature points for computing the local stress state. We employ the Mechanics of defect evolution library (MoDELib)\cite{pomechanics}, which implements phenomenological mobility laws to model dislocation dynamics~\cite{po2014recent}. DDD has traditionally been utilized to study dislocation-dislocation interactions, and dislocation reactions with other types of material defects such as second phases (Eshelby inclusions) or SF tetrahedra \cite{martinez2008dislocation, martinez2008dislocation2}. Many of these works were conducted using pure materials 
In this work, the foundation for adding material complexity into DDD is provided, including strengthening effects from long-range solid solution elastic effects, and short-range SF chemical effects. The framework is implemented and integrated to the MoDELib DDD package.

\subsection{Computation of Locally Varying Stacking Fault Energy}\label{sec:MethodSub2}

The SFE landscape in different Al-Mg solid solutions is computed using non-orthogonal MD simulation cells. The composition range is limited to 5\%-15\% at. Mg to avoid strong core effects, such as constriction and cross-slip of screw segments, which emerge at higher Mg concentrations~\cite{nohring2017dislocation}. 

An embedded-atom method (EAM) potential for Al-Mg~\cite{liu1998grain,liu1996eam,liu1997anisotropic} is used to compute the configurational energies. The samples are prepared by creating a fully periodic, pure Al system. Subsequently, Al atoms in the system are replaced by Mg atoms at random to create AlMg$_x$ alloy samples, with target concentration of 5, 10, and 15 at-\% Mg. The samples are constructed in a non-orthogonal coordinate system defined by $x = [\bar{1}\bar{1}0]$, $y = [\bar{1}01]$, and $z = [\bar{1}1\bar{1}]$. The atomic system size is fixed at 24,300 atoms, with a cross-sectional area of approximately 71.23 nm$^2$. Periodicity along the $z$-axis is broken by introducing vacuum layers above and below the simulation cell.

\begin{figure}[H]
    \centering
    \adjincludegraphics[,width=0.8\textwidth,trim={{.0\width} {.0\width} {.0\width} {.0\width}},clip]{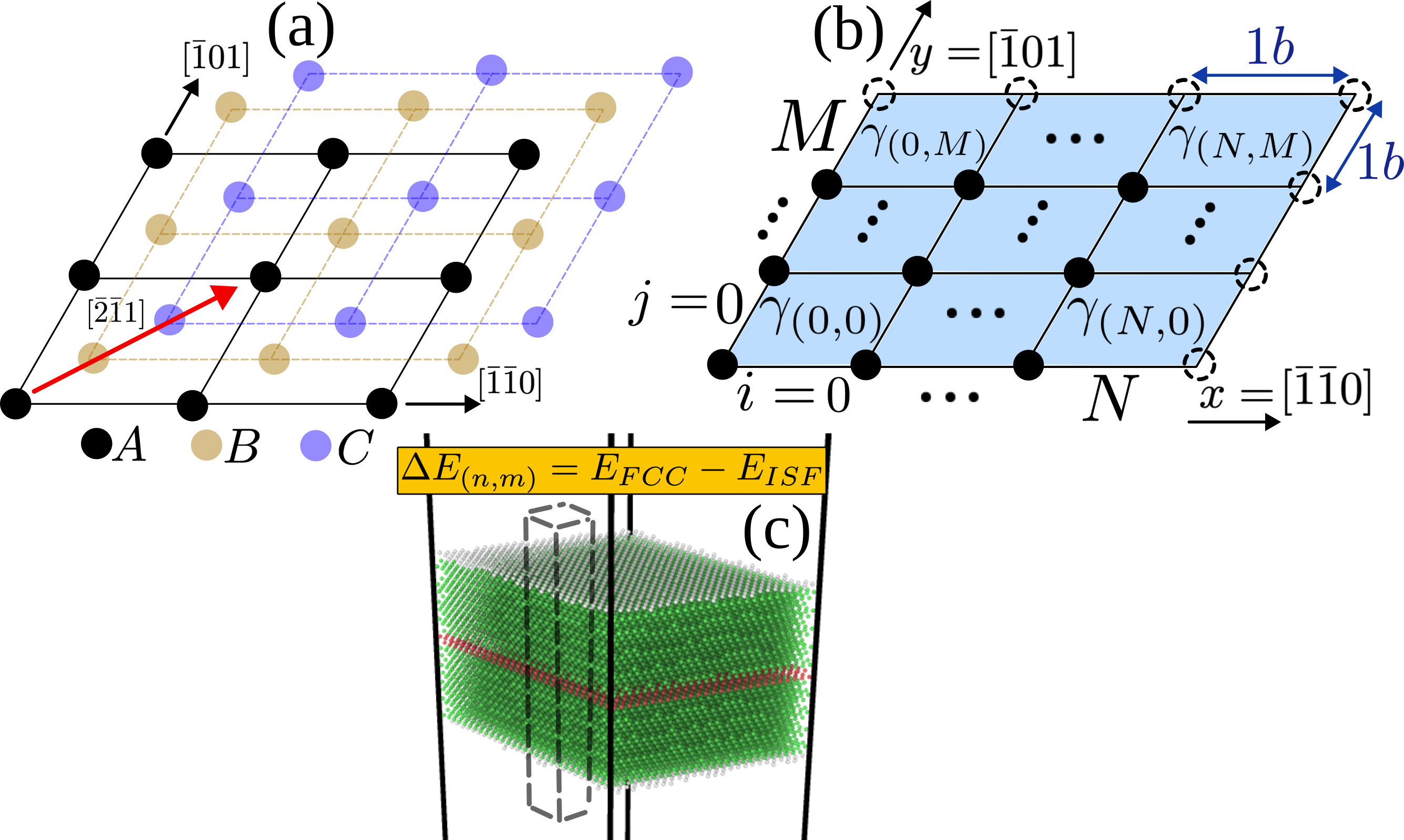}
     \caption{(a) Stacking sequence of the FCC structure captured by each grid cell. (b) Constructed spatial fault energy grid. Black circles represent A-type stacking atoms on both figures. The empty circles drawn with a dotted line represent the periodicity of the grid. Each grid cell in (b) is defined by four A-type stacking atoms. (c) Visualization of the ISFE grid cell. Red atoms in the center represent local hexagonal close-packed (HCP) stacking, while green atoms represent FCC stacking.}
    \label{fig: fault_energy_calc}
\end{figure}

Although the system contains a relatively small number of atoms, no significant finite-size effects are observed in larger systems tested during the sampling process. The crystal shown in Figure~\ref{fig: fault_energy_calc}c is divided into two regions: upper and lower crystals. A displacement of the top region along the $[\bar{2}\bar{1}1]$ direction—by half, one, and two times the partial Burgers vector—is applied to generate unstable stacking fault (USF), ISF, and maximum stacking fault (MSF) configurations, respectively. Molecular statics (MS) with the conjugate gradient method is used to find the energy minimum at zero stress of a given configuration. It is worth noting that the atomic motion perpendicular to the glide normal is restricted when the stacking positions are on either USF or MSF to capture the accurate unstable energy values. Although USF and MSF configurations are sampled, we focus on ISF, as it is most directly related to the glide behavior of extended dislocations.

To compute the statistical distribution of local ISF energies, 100 independently generated random alloy configurations were used.
To isolate the effect of solutes on measuring locally fluctuating ISFEs, the sample is regularly meshed in $x$ and $y$ with a grid spacing of one Burgers vector, as this represents the smallest displacement capable of representing SF area. Each atom is assigned to a cell according to their $x$, $y$ coordinates. The vertices of each cell are denoted as A atoms in Figure \ref{fig: fault_energy_calc}a, as atoms in A planes in the stacking sequence. For each grid cell, the ISF energy is calculated using Eq. \ref{eq:sfe}. 
\begin{equation}
    \gamma^{\text{ISF}}_{(n,m)} = \frac{E_{ISF}-E_{FCC}}{\mathcal{A}}
    \label{eq:sfe}
\end{equation}
$\gamma^{\text{ISF}}_{(n,m)}$ is computed by subtracting the energy within a cell before displacement from their energy after displacement, calculating the grid energies as the sum of atomic energies of the atoms belonging to cell $(n,m)$ dividing by the area of the plane $\mathcal{A}$. The resulting $\gamma^{\text{ISF}}_{(n,m)}$ is a scalar random field characterized by mean $\bar{\gamma}$ and standard deviation $\sigma_{\gamma}$.

\begin{figure}[H]
    \centering
    \adjincludegraphics[,width=0.8\textwidth,trim={{.0\width} {.0\width} {.0\width} {.0\width}},clip]{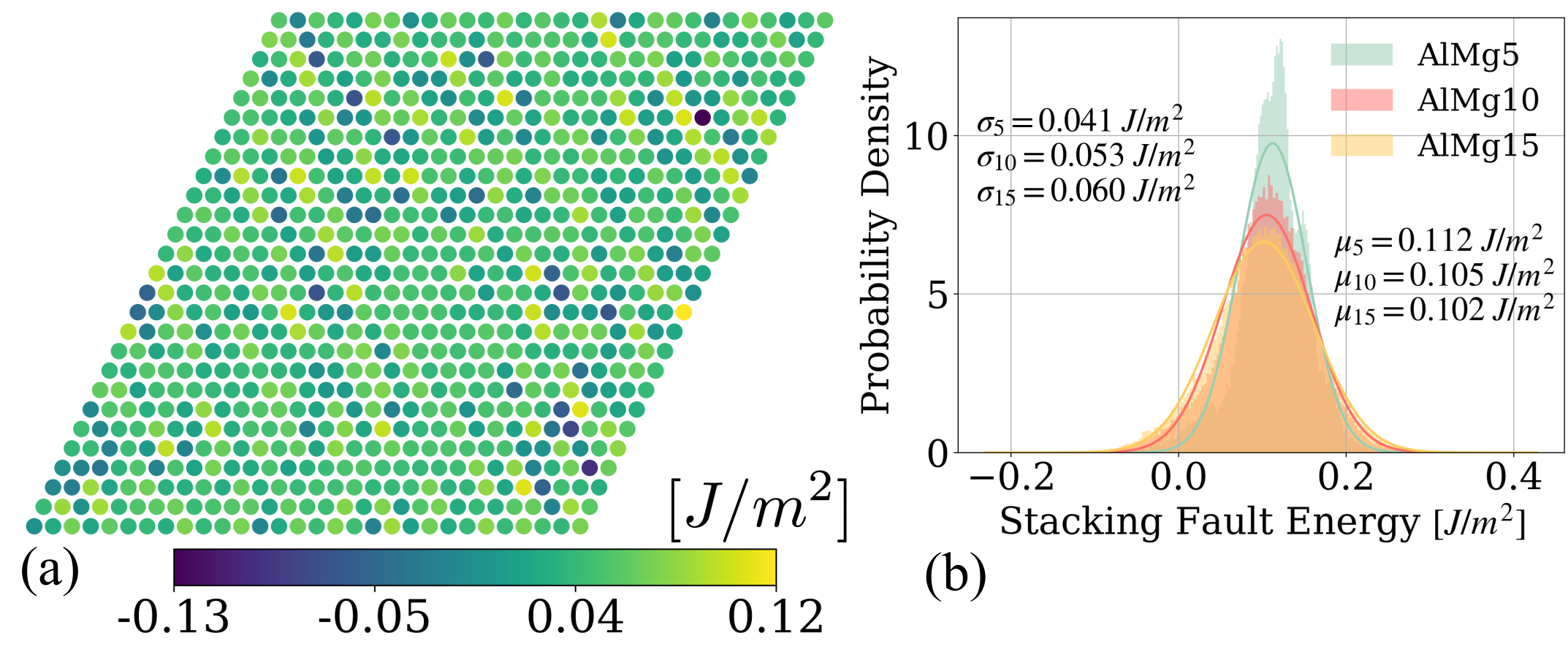}
    \caption{(a) $\gamma_{\text{ISF}}$ at each grid point in an AlMg-5at\%. (b) The $\gamma_{\text{ISF}}$ distribution depending on the Mg content.}
    \label{fig:SFEdot}
\end{figure}

Figure~\ref{fig:SFEdot}(a) shows the $M \times N$ scalar ISFE field constructed using the aforementioned scheme. To enhance visual clarity, each grid point is represented as a circular dot, reflecting the locally varying $\gamma^{\text{ISF}}_{(i,j)}$ values. A viridis color scheme is used for visualization, where dark dots indicate lower energies and bright dots correspond to higher energies. Figure~\ref{fig:SFEdot}(b) presents the distribution of $\gamma^{\text{ISF}}_{(i,j)}$ values as a function of alloy composition. We observe that the standard deviation increases with Mg content, due to larger configurational fluctuations, while the mean decreases.

\subsection{Spatial Autocorrelation of SFE}\label{sec:MethodSub3}






Once the locally varying fault energy scalar field is constructed, the spatial autocorrelation of the ISFE is computed using the following equation:

\begin{equation}
   R(n,m) = \left\langle \sum_{i=0}^{N} \sum_{j=0}^{M} \delta \gamma (i,j) \, \delta\gamma (i-n,j-m) \right\rangle
    \label{eq:correlation}
\end{equation}

\noindent where $N$ and $M$ are the number of grid points in the $x$ ($N$ = 30 points) and $y$ ($M$ = 27 points) directions, respectively; $n$ and $m$ are the lag parameters used to compute the correlation; and $i$ and $j$ denote the grid indices at each spatial location, with $\delta \gamma (i,j)=\gamma^{\text{ISF}}_{(i,j)}-\bar{\gamma}$. The angular brackets $\langle \cdot \rangle$ denote ensemble averaging over all valid positions. The computation is repeated across 100 independent random solid solution configurations and averaged to obtain statistically converged autocorrelation information, denoted as $\langle R \rangle_{100}$. Once the autocorrelation map is obtained, the complete spatial correlation map is constructed using the positional information of the non-orthogonal grid designed to capture the locally varying SFE. This process is repeated for the Mg contents investigated in this work.

\begin{figure}[H]
    \centering
    \adjincludegraphics[,width=0.9\textwidth,trim={{.0\width} {.0\width} {.0\width} {.0\width}},clip]{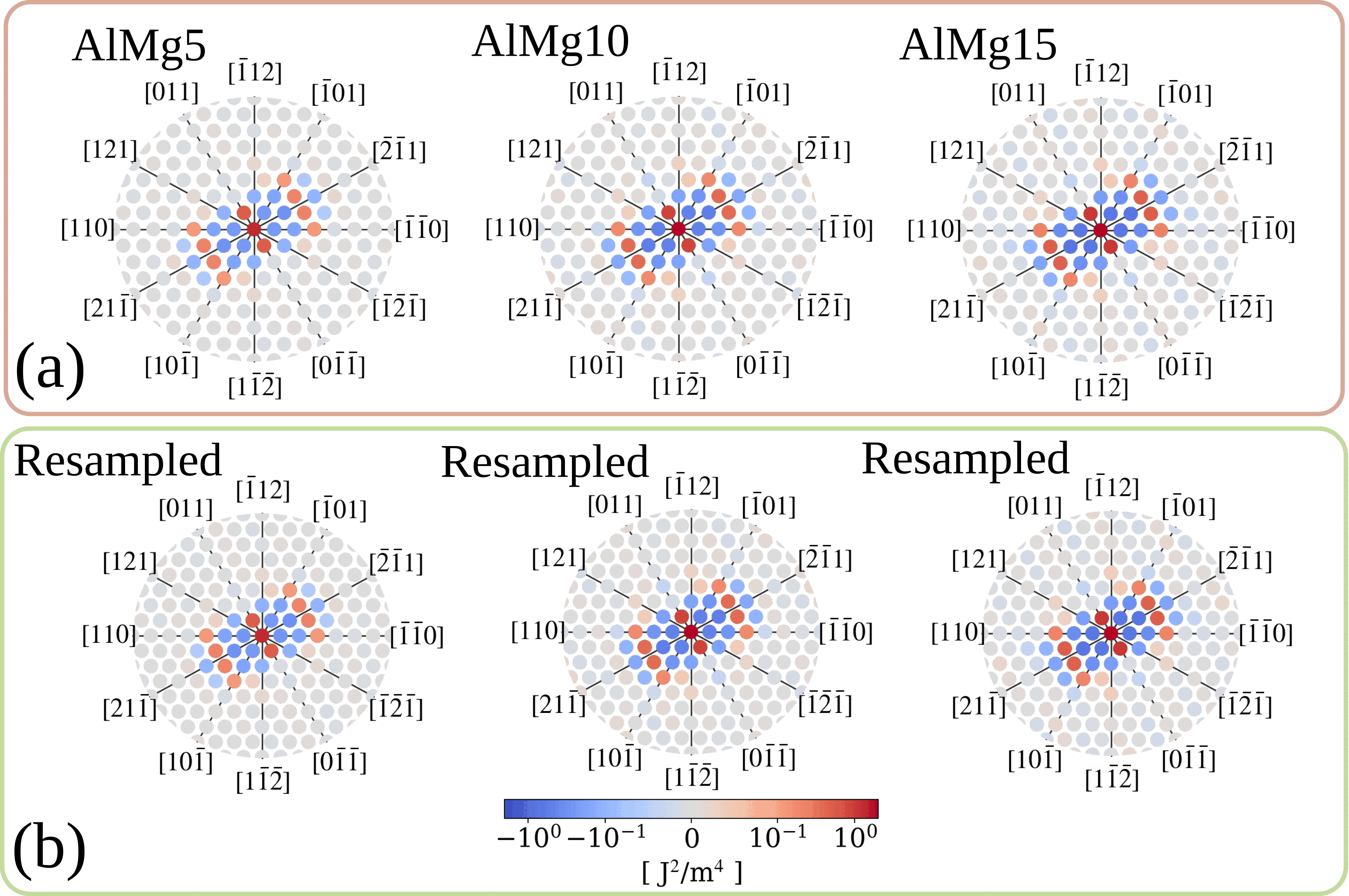}
    \caption{(a) Spatial correlations sampled from the non-orthogonal MD systems depending on Mg content with $[\bar{2}\bar{1}1]$ displacement direction. (b) Recomputed correlations from the sampled random spatial SF noise. The spatial correlation data is plotted in logarithmic scale and trimmed to a circular shape. The zone axis crystallographic directions are added for better visualization.}
    \label{fig:121Circle}
\end{figure}

\begin{figure}[H]
    \centering
    \adjincludegraphics[,width=1\textwidth,trim={{.0\width} {.0\width} {.0\width} {.0\width}},clip]{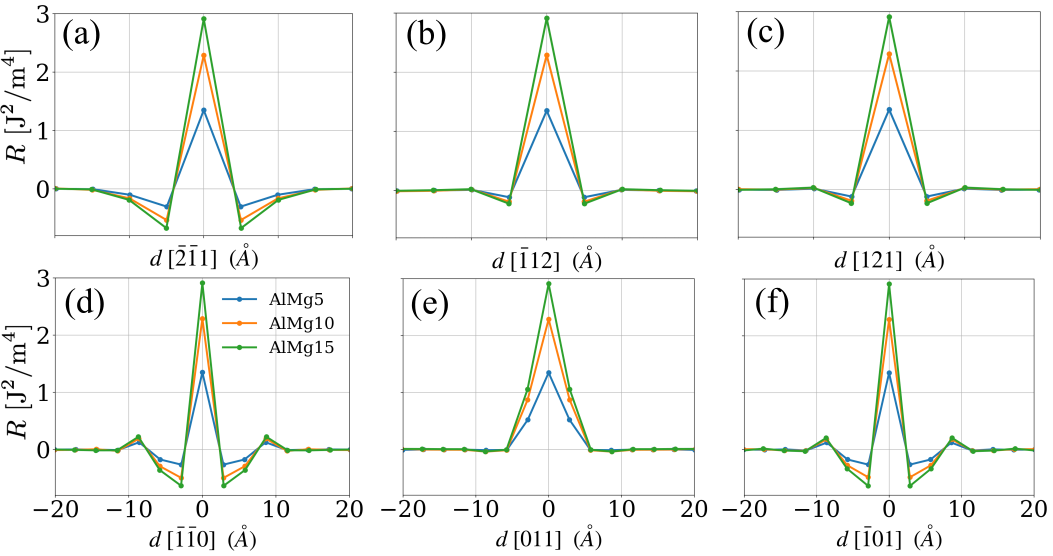}
    \caption{Spatial correlation data of $\Delta \gamma _{ISF}$ in different crystallographic direction. The correlations decay and converge to zero within around 10-15 \AA. The negative correlation trend becomes more pronounced with the presence of Mg atoms.}
    \label{fig:1Dauto}
\end{figure}

Figures \ref{fig:121Circle} and \ref{fig:1Dauto} show the un-normalized spatial autocorrelations computed with MD depending on the alloy composition. As Mg content increases, the magnitude of the correlations increases. Figure \ref{fig:121Circle} presents 2D color maps in a zone axis projection with explicit directions, while Fig. \ref{fig:1Dauto} displays a 1D projection of the autocorrelation values in different directions. We observe that the range of the correlations is about 10–15 \AA, equivalent to approximately 4 Burgers vectors. These distances are commonly referred to as characteristic correlation lengths~\cite{rida2022influence, patinet2013quantitative, park2003dynamics}. We also see that the correlations are highly anisotropic. The correlation along the displacement direction (Fig. \ref{fig:1Dauto}(a)) shows a change of sign with respect to distance, from positive close to $0$ to negative and then smoothly tending to $0$. The correlations computed in the other $\langle 112 \rangle$ directions, i.e. $[\bar{1}12]$ and $[\bar{1}\bar{2}\bar{1}]$, follow similar trends, with a crossover from positive to negative values (Figs. \ref{fig:1Dauto}(b)-(c)). In these last two cases, the range of correlations is slightly shorter than in the displacement direction ($[\bar{2}\bar{1}1]$) with weaker correlation values. Furthermore, in two of the $\langle 110 \rangle$ directions, $[\bar{1}\bar{1}0]$ and $[\bar{1}01]$, the correlations display more complex features, going from positive to negative to positive before tending to $0$ (Figs. \ref{fig:1Dauto}(d) and (f)). These two directions are symmetric with respect to the displacement direction and show similar features and values. On the other hand, correlations in the $[011]$ direction (Fig.~\ref{fig:1Dauto}(e)) show mainly positive values with a slightly negative point before tending to $0$.

\begin{figure}[H]
    \centering
    \adjincludegraphics[,width=0.7\textwidth,trim={{.0\width} {.0\width} {.0\width} {.0\width}},clip]{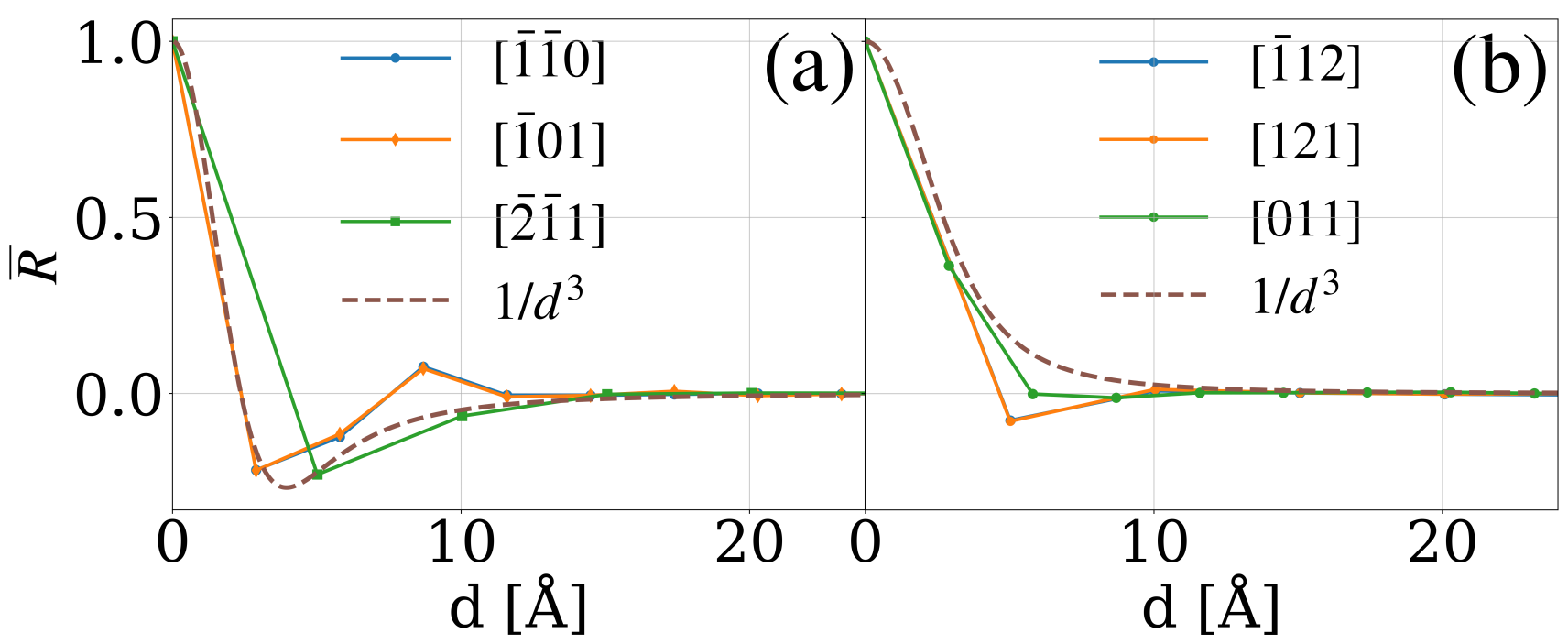}
    \caption{SF auto-correlation comparison showing (a) longitudinal and (b) transverse long-range trends observed in Ref.~\cite{geslin2021microelasticity,geslin2021microelasticity2}. Correlation magnitudes are normalized using AlMg15 reference data.}
    \label{fig:long_range_comp}
\end{figure}

The autocorrelations observed in our local SF energy fluctuations are compared with stress correlation expressions derived from Geslin et al.'s studies~\cite{geslin2021microelasticity,geslin2021microelasticity2} in Figure \ref{fig:long_range_comp}. Directions exhibiting notable negative correlations ($[\bar{1}\bar{1}0]$, $[\bar{1}01]$, and $[\bar{2}\bar{1}1]$) are compared to the longitudinal stress correlation ($\Sigma_L$) expression, while directions showing positive or marginally negative correlations ($[\bar{1}12]$, $[121]$, and $[011]$) are compared to the transverse stress correlation ($\Sigma_T$) expression. The corresponding stress correlation equations are provided below.

\begin{equation}
    \begin{aligned}
    &\Sigma_L = -30\left(\frac{a}{d}\right)^3 \left[\sqrt{\pi}\left(1-\frac{12a^2}{d^2}\right)erf\left(\frac{d}{2a}\right)+\frac{a}{d}\left(12+\frac{d^2}{a^2}\right)exp\left(-\frac{d^2}{4a^2}\right)\right] \\
    &\Sigma_T = 15\left(\frac{a}{d}\right)^3 \left[\sqrt{\pi}\left(1-\frac{6a^2}{d^2}\right)erf\left(\frac{d}{2a}\right)+\frac{6a}{d}exp\left(-\frac{d^2}{4a^2}\right)\right]
    \end{aligned}
    \label{eq:SSCorr}
\end{equation}

Interestingly, the SF correlations along the crystallographic directions exhibit a consistent long-range decay of $\approx 1/d^3$. A previous study on local stress fluctuations revealed that the $1/d^3$ power-law tails are the signature of normal stress fluctuations in disordered isotropic solids~\cite{lemaitre2018stress}. This indicates that our fluctuating SFEs exhibit similar characteristics with respect to stress fluctuations. However, the expressions shown in Eqs.~\ref{eq:SSCorr} are derived for stresses, and their use here is just a comparative exercise. 



The correlations sampled from each alloy system in MD simulations are used to generate correlated structural noise fields in our mesoscale DDD model. A square $N \times M$ grid is constructed on the dislocation glide plane with a spacing of one Burgers vector, as described in Section~\ref{sec:MethodSub2}. Randomly fluctuating ISFE fields with correct spatial correlation statistics are then generated using the random grating method~\cite{maradudin1990enhanced}, a widely used technique for sampling correlated random fields. The details of the sampling method can be found in ~\ref{sec:app_correlated_noise}. In short, the random field is first computed by convolving the Gaussian random variables with the square root of the sampled correlation spectrum in Fourier space. Specifically, we compute:

\begin{equation}
   \widehat{\psi}(k_x,k_y)=\sqrt{\widehat{R} (k_x,k_y)}\frac{\widehat{\mathcal{Q}}+i\widehat{\mathcal{P}}}{\sqrt{2}}
    \label{eq:realSpaceConvolution}
\end{equation}


\noindent where $\widehat{\psi}(k_x,k_y)$ represents the random amplitude based on ISF in Fourier space, and $\widehat{R} (k_x,k_y)$ is the Fourier transform of the correlations computed with MD. 
Notably $\widehat{\psi}(k_x,k_y)$ embeds noise amplitudes calibrated to the variance of spatial autocorrelation trends in AlMg5, AlMg10, and AlMg15 alloys, as the correlations retain their absolute magnitudes unlike conventional normalized approaches. 
 Once anti-transformed, $\psi (\bm{x})$ holds the fluctuations of the ISFE on the grid with the same correlations as obtained in MD.

The distributions and correlations that we have obtained from our MD sample are converged, i.e., there is no change if we increase the size of the atomistic box. To conserve the statistical features of the distribution in DDD, the correlation to be sampled is padded with zeros to make it commensurate with the size of the glide plane at the mesoscale. $\psi(\bm{x})$ is normalized to match that of the unpadded noise field, as detailed in ~\ref{sec:std_pad_correction}.

\begin{figure}[H]
    \centering
    \adjincludegraphics[,width=0.8\textwidth,trim={{.0\width} {.0\width} {.0\width} {.0\width}},clip]{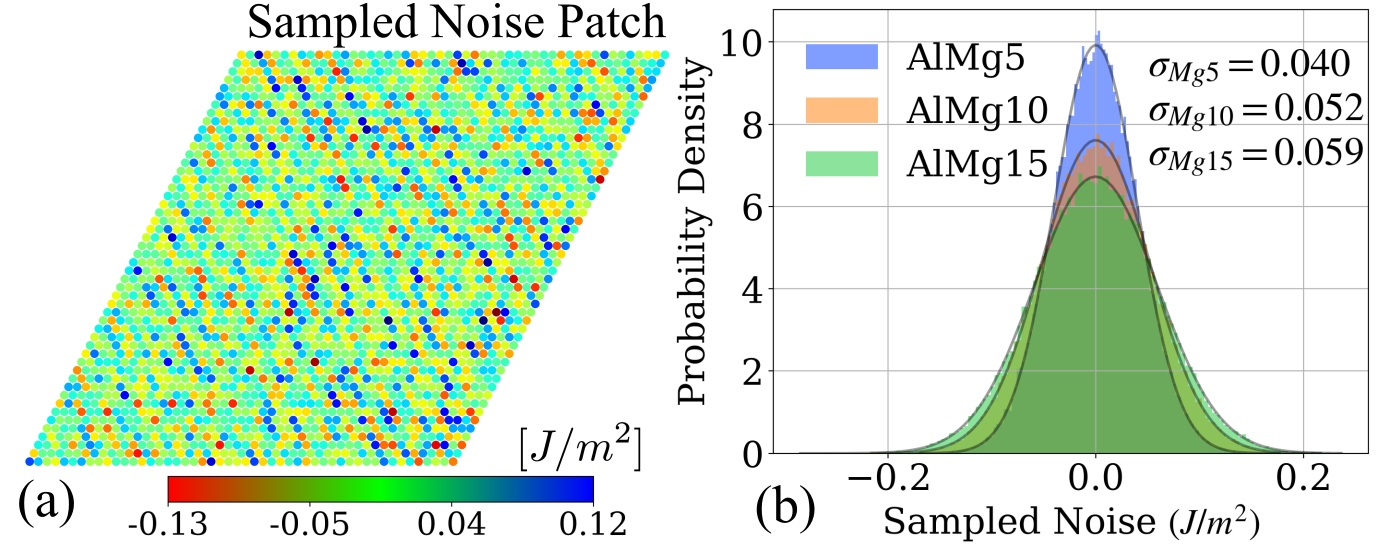}
    \caption{A single realization of an AlMg-5at\% random noise surface sampled with ISF correlation is depicted in Figure (a) as an example. A thousand realizations of the sampled correlated random noises are plotted in Figure (b). The statistical properties of the MD data are reproduced by the random surface model.}
    \label{fig:zeroPad}
\end{figure}

To verify the correctness of our zero-padding scheme, the ISF is sampled within our DDD implementation, and its autocorrelation is computed by averaging over 1,000 realizations. Results are shown in Figs.~\ref{fig:121Circle} and \ref{fig:zeroPad}, where we observe that the autocorrelation is preserved, indicating that the statistical features of the fluctuations are accurately reproduced. 


This field is applied to the glide plane in DDD as a 2-D orthogonal grid. Then, a bilinear interpolation algorithm is employed to sample the noise contribution on the SFE in our mesoscale simulation. Specifically, the SFE contributions arising from the sampled noise are interpolated based on the locations of Gauss quadrature points (GPs) on the non-orthogonal grid. 

\begin{figure}[H]
    \centering
    \adjincludegraphics[,width=0.8\textwidth,trim={{.0\width} {.0\width} {.0\width} {.0\width}},clip]{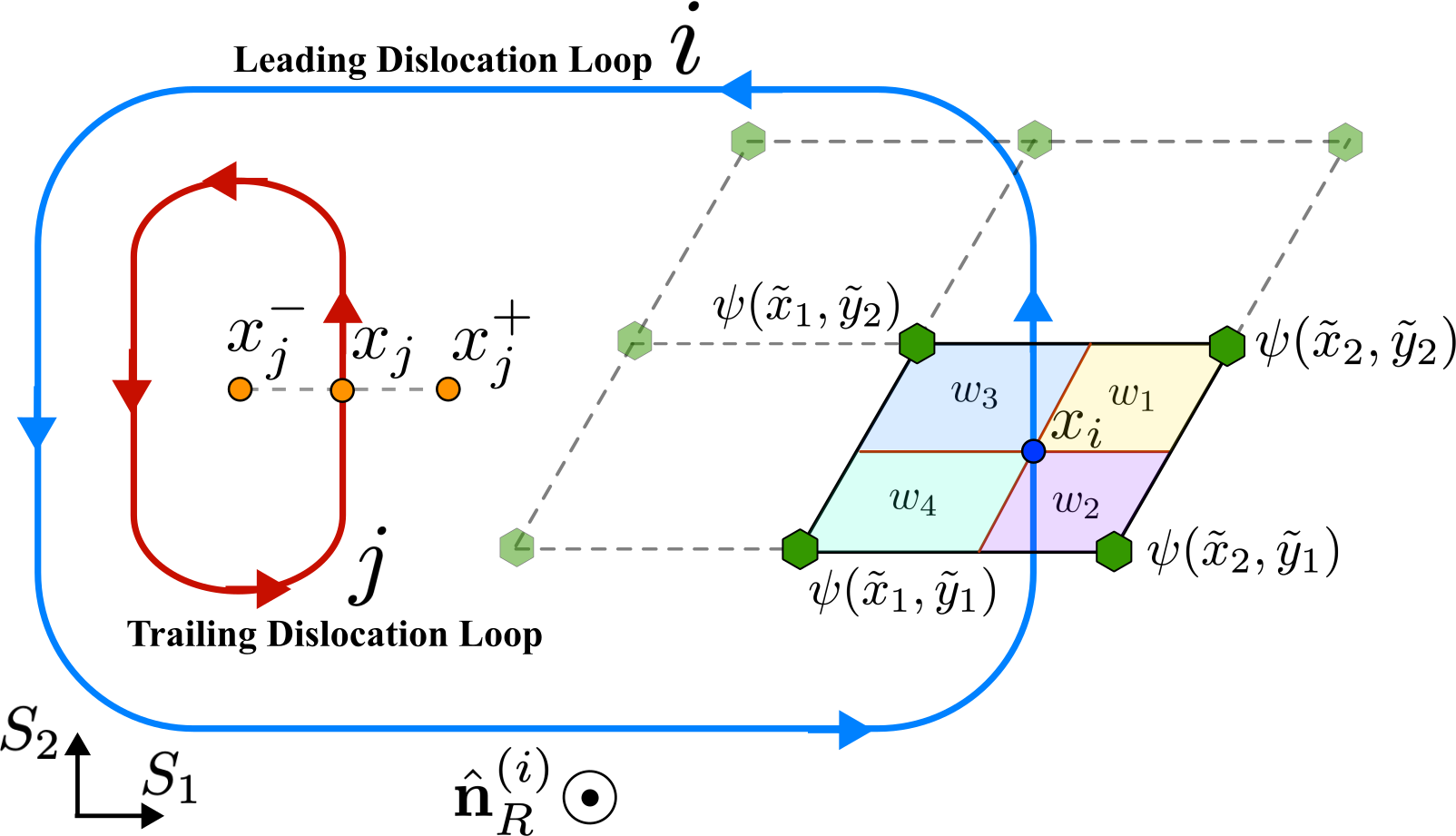}
    \caption{Visualization of bilinear interpolation based on GPs $x_i$ and $x_j$. The blue and red lines represent leading and trailing dislocation lines, respectively. The contribution from fluctuating SFE values, $\psi(\bm{x})$, at each green grid point is weighted by $w_i$, computed from the enclosed parallelogram area. Only the noise field for the leading dislocation line is drawn for figure clarity.}
    \label{fig:bilinear}
\end{figure}

The contribution of the ISF to the force on a GP along the dislocation line is computed following the description in Ref.~\cite{po2014recent,li2021cross,liu2025atomistically}. The gliding partial dislocations on a slip plane induce the change in the so-called misfit energy. This generates a configurational force which acts as part of the driving force for the dislocations. The spatially fluctuating field, $\psi(\bm{x})$, consists of discrete data with finite resolution, as shown in Figure~\ref{fig:bilinear}. The SFE density on the plane at point $\bm{x}$ relates to the slip vector $\bm{s}(\bm{x})$. The slip vector is defined by:

\begin{equation}
    \bm{s}(\bm{x}) = - \sum_i W_i(\bm{x}) \bm{b}_i 
    \label{eq:displacement_field}
\end{equation}

\noindent where $W_i(\bm{x})$ is a winding number borrowed from complex analysis~\cite{krantz1999handbook} and $\bm{b}$ is the Burgers vector. Specifically, $W_i(\bm{x})$ represents the winding number at position $\bm{x}$ around dislocation $i$. The winding number is an integer and measures how many times the loop winds around point $\bm{x}$. We define the vector $\bm{N} = \bm{s}_1 \times \bm{s}_2$, where $\bm{s}_1$ and $\bm{s}_2$ are Cartesian unit vectors along the $\langle110\rangle$ and $\langle112\rangle$ crystallographic directions, respectively. $\bm{n}_R$ represents the normal to the dislocation loop plane and the relation between $\bm{N}$ and $\textbf{n}_R$ determines the directionality of winding:

\begin{equation}
\begin{aligned}
    &\bm{N} \cdot \textbf{n}_R = 1 \ \text{then} \ W(\bm{x}) = 1  \,, \text{clockwise} \\
    &\bm{N} \cdot \textbf{n}_R = -1 \ \text{then} \ W(\bm{x}) = -1 \,, \text{counterclockwise} \\
    \label{eq:}
\end{aligned}
\end{equation}

The points $\bm{x}_i^+ = \bm{x}_i + \varepsilon \bm{d}_i^+$ and $\bm{x}_i^- = \bm{x}_i - \varepsilon \bm{d}_i^-$ define the positions immediately adjacent to the dislocation line. In our current scheme, the SFE is both a function of the slip vector and the position of the GP due to the spatial fluctuations. Namely, the function $\gamma(\bm{x},\bm{s}(\bm{x}))$ represents the SFE of the glide plane at point $\bm{x}$. 
\begin{equation}
    \gamma(\bm{x},\bm{s}(\bm{x})) = \bar{\gamma}(\bm{s}(\bm{x})) +\psi(\bm{x})
    \label{eq:sf_energy}
\end{equation}

\subsection{Mesoscale Critical Resolved Shear Stress (CRSS)}\label{sec:MethodSub4}

To determine the critical stress required for dislocations to overcome obstacles introduced by both solid solution (SS) and SF noise effects, a series of DDD simulations are conducted in periodic simulation cells containing dislocation dipoles. \HL{Following the eigenstrain-based framework of Geslin et al.~\cite{geslin2021microelasticity,geslin2021microelasticity2}, we compute the correlated SS noise fields on an orthogonal grid with a 0.05 nm spacing.} To account for periodicity, one periodic image is used in each dimension to reduce computational cost, as our tests revealed that the depinning transitions are not significantly affected by the number of images. A mesh with the crystal orientations $x[011]$, $y[\bar{2} \bar{1} 1]$, and $z[\bar{1} 1 \bar{1}]$ was created. 


To minimize the effect of the dislocation dipole interaction, each glide plane is separated by 2,500 adjacent lattice planes for a total length in the direction normal to the glide planes of $5900b$. The gliding direction length and height along the glide plane are fixed at $1020b$ and $5900b$, respectively.
Moreover, the grid spacing for sampled noise is fixed at $1b$ along the $x$- and $y$-axes of the glide plane, as this represents the highest resolution of local SF fluctuations achievable (see Sec.~\ref{sec:MethodSub2}). The grid sizes are set commensurate with simulation box sizes, which are determined based on the corresponding dislocation lengths.

\begin{table}[htbp]
\centering
\begin{tabular}{lllll}
\hline
Type & Symbol & Value & Unit & Ref. \\
\hline\hline
Line tension paramter & $\Gamma_{pd}$ & 1.50 &  & \\
Non singular parameter& $a_{NS, pd}$ & 0.90 &  & \\
Burgers vector & $b$ & 2.86 & \AA & \\
Drag coefficient (edge)& $B_{0,e}$ & 13.4 &$\mu\text{Pa}\cdot\text{s}$ & \cite{geslin2018thermal} \\
Drag coefficient (screw)& $B_{0,s}$ & 13.4 & $\mu\text{Pa}\cdot\text{s}$ & \cite{geslin2018thermal} \\
Poission ratio & $\nu_{Mg5}$ & 0.341 &  & \\
 & $\nu_{Mg10}$ & 0.343 &  & \\
 & $\nu_{Mg15}$ & 0.345 &  & \\
Shear modulus & $\mu_{Mg5}$ & 28.6 & GPa & \\
 & $\mu_{Mg10}$ & 26.8 & GPa & \\
 & $\mu_{Mg15}$ & 25.0 & GPa & \\
Mean ISFE & $\gamma_{ISF,Mg5}$ & 0.112 & $\text{J/m}^2$ & This work \\
& $\gamma_{ISF,Mg10}$ & 0.105 & $\text{J/m}^2$ & This work \\
& $\gamma_{ISF,Mg15}$ & 0.102 & $\text{J/m}^2$ & This work \\
\hline\hline
\end{tabular}
\caption{DDD parameters used in this work}
\label{Table:dddparams_combined}
\end{table}

The nodal density on each dislocation line is set according to the dislocation character. The density must be high enough to accurately capture the effect of the noise but cannot exceed a threshold due to a phenomenon called the ultraviolet catastrophe~\cite{boleininger2020ultraviolet}. This instability arises from numerical artifacts that can be avoided by choosing appropriate line tension parameters. 


The line tension model used for this investigation accounts for contributions from linear elasticity and core energy. To avoid the singularity arising from Voletrra's model in segment interactions, we adopt the non singular framework \cite{cai2006non}. The core (self) energy contribution is included via Blin's self-energy expression \cite{blin1955energie}, which is given in Eq. \ref{eq:energy_density}.


\begin{equation}
    e(\hat{\bm{T}}) = \alpha \frac{\mu}{4\pi(1-\nu)}\left(b^2-\nu(\hat{\bm{T}}\cdot\bm{b})^2\right)
    \label{eq:energy_density}
\end{equation}

\noindent 

where $\hat{\bm{T}}$ is the normalized line tangent, $\bm{b}$ is the Burgers vector (with magnitude $b$), and $\alpha$ is the line-tension scaling factor. In an isotropic medium, the line-tension force on each segment is computed as the energy density multiplied by the curvature norm, $|\bm{\kappa}|$. We carefully finetune $\alpha$ to mitigate numerical instability, ultimately adopting a value of 1.5—which yields stable partial configurations that do not collapse under the high SFE of Al-Mg. The nodal density for edge dislocations is set to approximately 0.7 nodes/$b$, and for screw dislocations, to approximately 0.8 nodes/$b$. These values maintain system stability while providing sufficient flexibility for dislocations to accommodate the energy landscape. The GP density is set such that the distance between points is $\approx$ 0.2$b$, ensuring that fluctuations as small as one Burgers vector can be resolved. All simulations use a fixed time integration method with a sufficiently small time step ($dt = 0.06 \ [b/c_s]$, where $c_s$ denotes the shear wave speed) to accurately capture noise effects on gliding motion and avoid instabilities.

 
\begin{figure}[H]
    \centering
    \adjincludegraphics[,width=0.7\textwidth,trim={{.0\width} {.0\width} {.0\width} {.0\width}},clip]{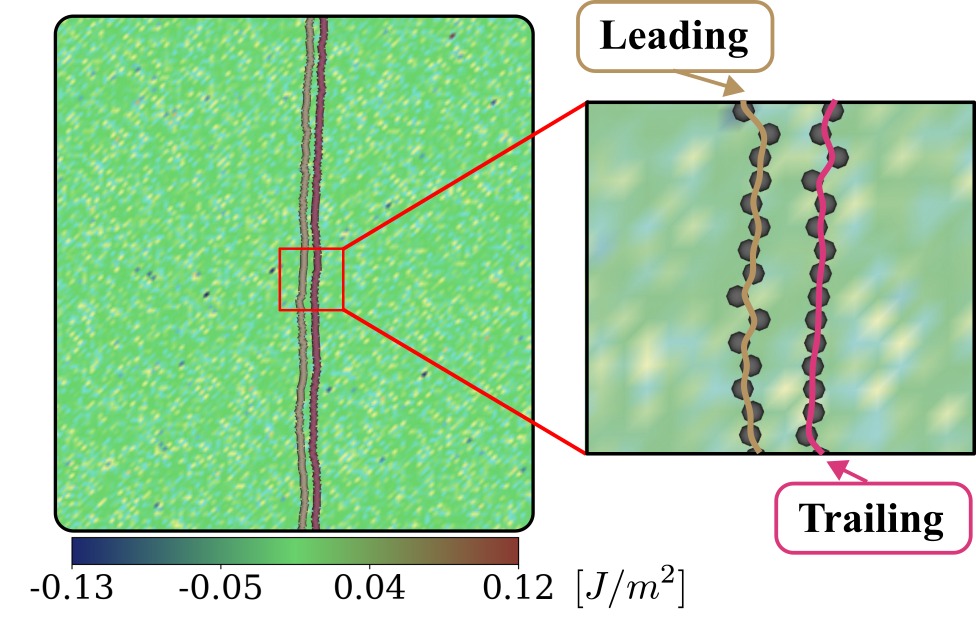}
    \caption{Gliding partial dislocations in an AlMg-5at\% correlated SF noise surface. Each dislocation line comforms to the noise to find the minimum energy state.}
    \label{fig:equiISFnoise}
\end{figure}
 
 All simulations are initially relaxed for 500 time steps to ensure they conform to an equilibrium separation distance. As shown in Fig.~\ref{fig:equiISFnoise}, the shape of the dislocations changes under the noise field to minimize the energy. 
 A constant external shear stress is then applied to drive dislocation glide. Each node is constrained to move on the glide plane in the direction perpendicular to the initial line direction, i.e., in the $x$ and $y$ directions for edge and screw dislocations, respectively, to avoid topological issues. 
 
 


The calculation of the CRSS, $\tau_{CRSS}$, is based on the rate of plastic distortion $\dot\beta^p_{ij}$ induced by gliding dislocations. Due to statistical ambiguity in the motion of a dislocation segment within a correlated stochastic noise field, we determine the CRSS by establishing upper and lower bounds ($\tau^+_{CRSS}$ and $\tau^-_{CRSS}$) found through the bisection algorithm described in Sec. \ref{sec:app_bisection}. The final CRSS is reported as $(\tau^+_{CRSS} + \tau^-_{CRSS})/2$.

Moreover, it is well documented that the depinning transition in random obstacles shows strong finite-size effects in terms of CRSS; the longer the dislocation line, the lower the estimated CRSS~\cite{bako2008dislocation,nogaret2006finite,vaid2022pinning}. Hence, the finite-size effect on CRSS is accounted for by controlling the glide distance observed before determining the CRSS. Namely, the time span over which we observe the dislocation evolution under applied stress is increased proportional to the length of the dislocation, as will be discussed in detail in Sec.~\ref{sec:ResultSub2}.


Due to the computational cost of DDD simulations, the stress resolution is limited to around 5-6 MPa. Each reported $\tau_{CRSS}$ value is averaged over 20 different noise configurations. The dislocation lengths tested for partial dislocations are 8.3, 18, 31.2, 42.9, 61.5, 70.1, 80.1 nm, respectively. All DDD simulations are conducted using stress-controlled dislocation evolution, with constant stress values applied throughout the simulation. The reported $\tau_{CRSS}$ is compared with the numerical results and predictions made with the models in Ref.~\cite{rida2022influence} for the case of correlated ISF noise and is further expanded to include the combined contributions from the ISF noise and solid solution noise. The parameters used in this work are described in Table \ref{Table:dddparams_combined}.


%% file: sections/results.tex
\subsection{Dislocation Length Convergence of the CRSS}\label{sec:ResultSub2}

It is known that the CRSS required for a dislocation to move depends on the dislocation length when the dislocation must overcome distributed obstacles to continue its glide. Under the line tension approximation, there is a characteristic length that marks the transition from a weakly pinned short dislocation to a strongly pinned long dislocation in disordered solid solutions. The exact definition of Larkin's length (beyond which the CRSS does not depend on length any longer) is still under debate~\cite{zapperi2001depinning, zaiser2002dislocation, zhai2019properties, rodney2024does}. 







\begin{figure}[H]
    \centering
    \adjincludegraphics[,width=0.84\textwidth,trim={{.0\width} {.0\width} {.0\width} {.0\width}},clip]{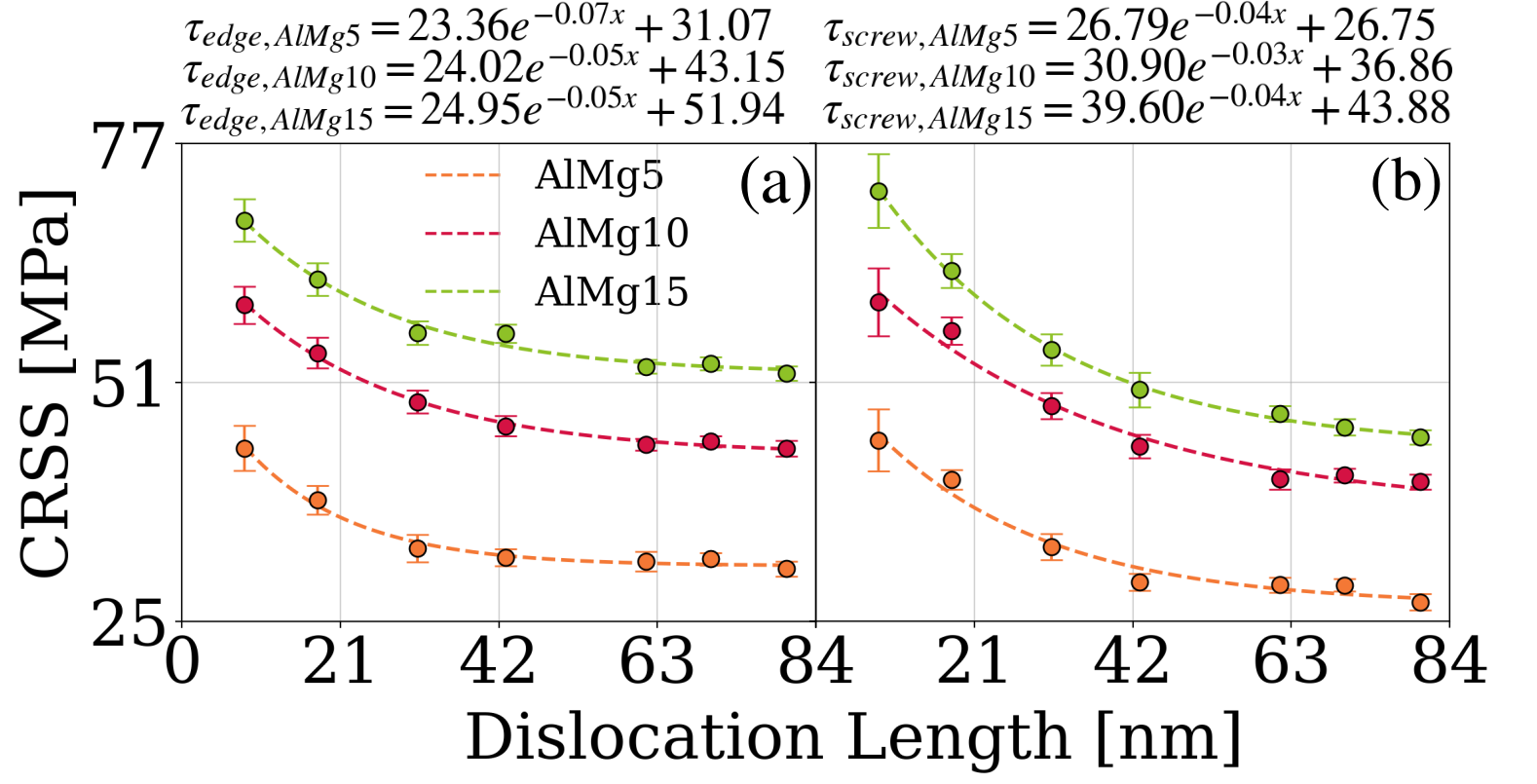}
    \caption{CRSS convergence as a function of dislocation lengths with correlated ISF noise and dislocation character (a) edge, and (b) screw. The dashed line represent the exponential fit shown in the figure.}
    \label{fig: indconv1}
\end{figure}

Figure \ref{fig: indconv1} shows the results for the CRSS depending on dislocation length and alloy composition, where we see that the CRSS decreases with dislocation length. A persistent size effect is observed even after the glide distance is varied in proportion to the dislocation length to control the probability of finding a weak obstacle. The CRSS increases as the Mg content increases for all lengths. The CRSS for both edge and screw seems to converge at a length of around 70-80 nm. 

\begin{figure}[H]
    \centering
    \adjincludegraphics[,width=0.7\textwidth,trim={{.0\width} {.0\width} {.0\width} {.0\width}},clip]{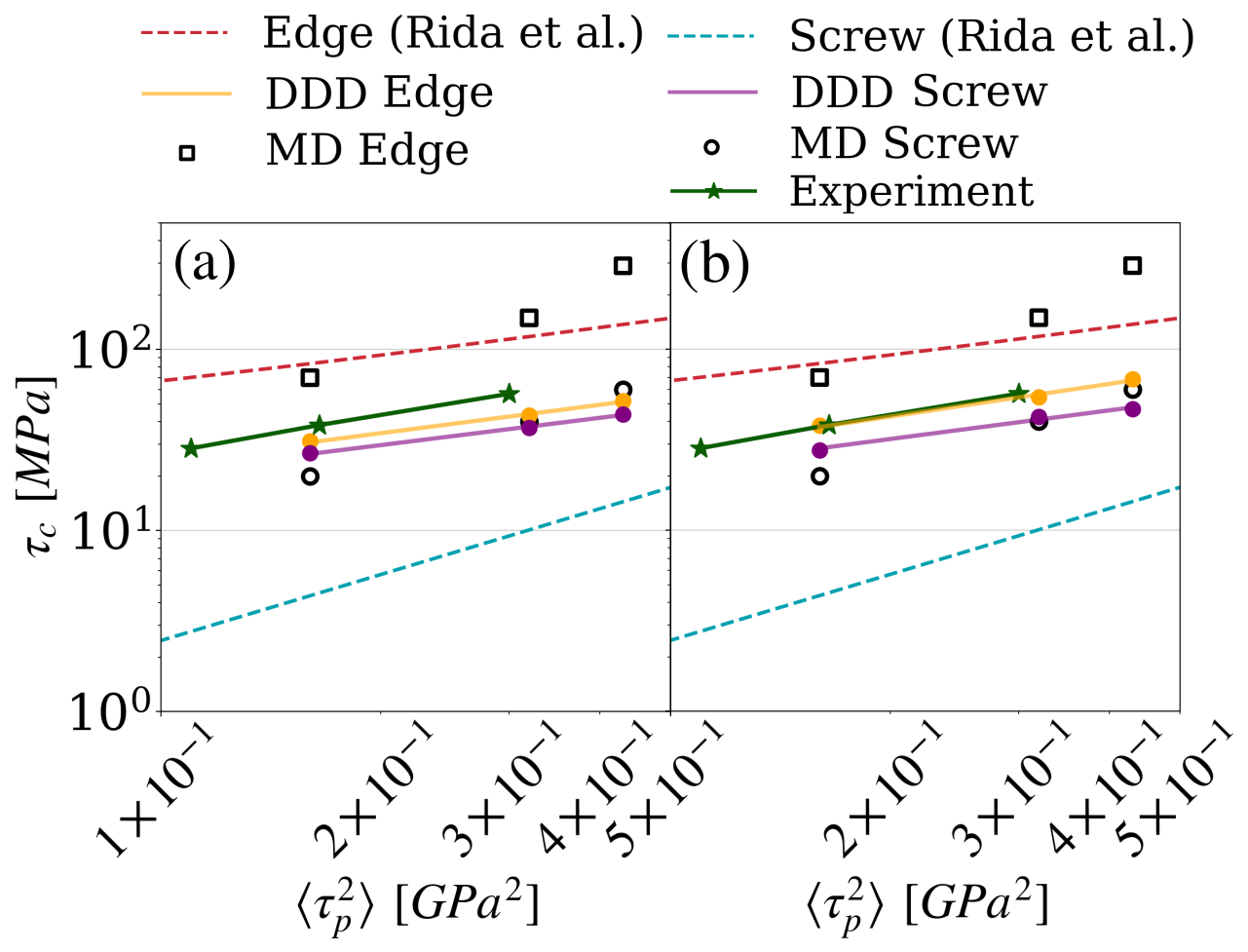}
    \caption{Comparison of CRSS data fitted to infinitely long dislocations with the model by Rida et al.~\cite{rida2022influence} (dashed lines). (a) Correlated stacking fault noise, (b) uncorrelated stacking fault noise.}
    \label{fig: infind_sf_u}
\end{figure}

Figure~\ref{fig: infind_sf_u} presents the CRSS results obtained from our DDD approach. For comparison, MD simulation data and the SS noise model developed by Rida et al.~\cite{rida2022influence} are also included. In addition, experimental values recorded at 1.6 K are plotted, as these are most relevant to our athermal simulation results~\cite{podkuyko1978peculiar}.
Note that the values are in a log-log scale. We clearly see that the ISF noise induces a non-zero CRSS in both screw and edge dislocations. The values due to the ISF noise lay in between the CRSS from the SS noise for screw and edge dislocations as calculated in Ref. \cite{rida2022influence}. Moreover, the values obtained from the simulations with ISF noise are close to the observed experimental values (green stars) and to the CRSS computed from MD simulations for screw dislocations (empty circles). The values of the CRSS induced by the solid solution on the edge dislocation from Rida's approach (dashed red line) and MD (empty squares) are significantly higher than those obtained with the ISF noise.


\begin{table}[]
\centering
\begin{tabular}{@{}lll@{}}
\toprule
Noise Type      & Character & Exponent \\ \midrule
Correlated SF   & Edge      & 0.46     \\
Correlated SF   & Screw     & 0.52     \\
Uncorrelated SF & Edge      & 0.59     \\
Uncorrelated SF & Screw     & 0.61     \\ \bottomrule
\end{tabular}
\caption{Power law fit exponents from DDD simulations. The exponents are fitted based on DDD results for AlMg5, AlMg10, and AlMg15-at.\% demonstrated in this work.}
\label{table: sf_exp_fit}
\end{table}

In general, it appears clear that ISF noise alone cannot account for the difference in CRSS values for screw and edge characters calculated with MD or from the SS model. Further discussion on the combined effects of ISF and SS noises is provided in Section \ref{sec:DiscussSub2}.


Next, we superimpose straight-line fits to the log-log data (Figure \ref{fig: infind_sf_u}). The extracted exponent values (Table \ref{table: sf_exp_fit}) are similar for correlated and uncorrelated cases. These values are moderately lower than the analytical values derived in our previous publication \cite{liu2025atomistically}, where a force-based sinusoidal model predicted exponents of 2/3 and 1 for uncorrelated and correlated noises, respectively. Note that for the edge and screw characters studied here, the correlations are weak, hence, we would expect exponents close to 2/3. We caution, however, that the present dataset may be insufficient to support a definitive conclusion. Furthermore, the correlations as a function of orientation are more complex than in our previous work \cite{liu2025atomistically}.



\subsection{Dislocation Character Dependence on CRSS}

\begin{figure}[H]
    \centering
    \adjincludegraphics[,width=0.8\textwidth,trim={{.0\width} {.0\width} {.0\width} {.0\width}},clip]{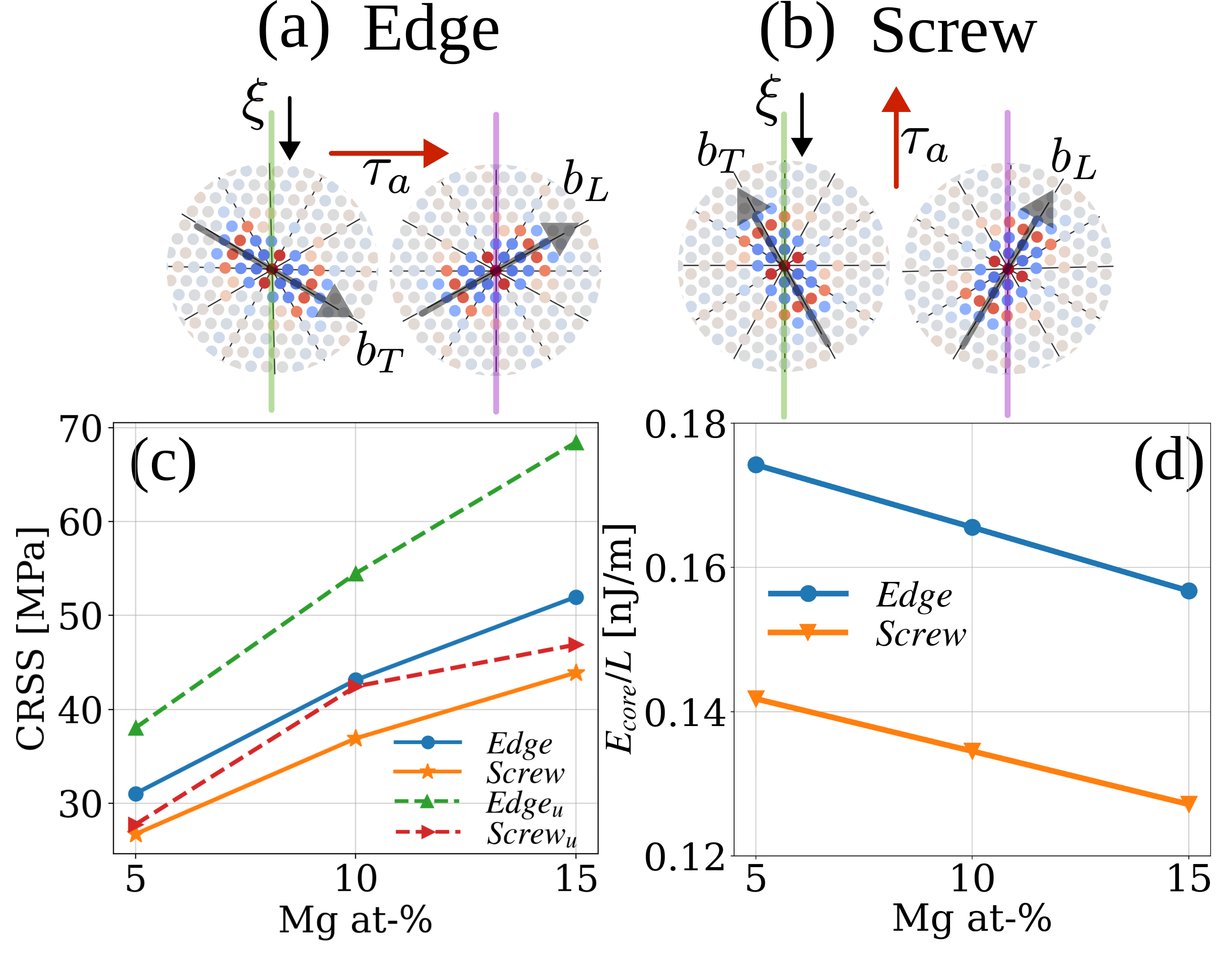}
    \caption{Spatial correlation of (a) edge and (b) screw partial characters; (c) CRSS for each character under correlated and uncorrelated SF noise (where the 'u' subscript denotes uncorrelated noise); and (d) core energy density ($E_{core}/L$) sum of partial pairs.}
    \label{fig:char_depend}
\end{figure}


Here, we examine the dependence of the dislocation character on CRSS in more detail. Partial pairs forming combined edge and screw character with a length of 61.5 nm are selected for this study. The ISF correlations observed along the tangent of each dislocation line ($\xi$) and Burgers vector ($b$) are shown in Figures \ref{fig:char_depend}(a) and (b). The CRSS of each configuration is found using the same bisection search algorithm described in Section \ref{sec:MethodSub4}. This procedure was repeated for 20 different ISF noise configurations and ensemble averaged over all realizations. Then, the data are fitted to the exponential functions shown in Figure \ref{fig: indconv1} so that the recorded data represent the CRSS values for an infinite dislocation length. When CRSS values are compared between uncorrelated and correlated ISF noise, the effect of correlation on the depinning transition stress is clear. The CRSS for both edge and screw characters under correlated SF noise is lower than under uncorrelated noise (see Figure \ref{fig:char_depend}(c)). This behavior reflects the negative correlation observed in our AlMg system and aligns with results reported in Refs. \cite{rida2022influence,liu2025atomistically}. 

\begin{figure}[H]
    \centering
    \adjincludegraphics[,width=0.6\textwidth,trim={{.0\width} {.0\width} {.0\width} {.0\width}},clip]{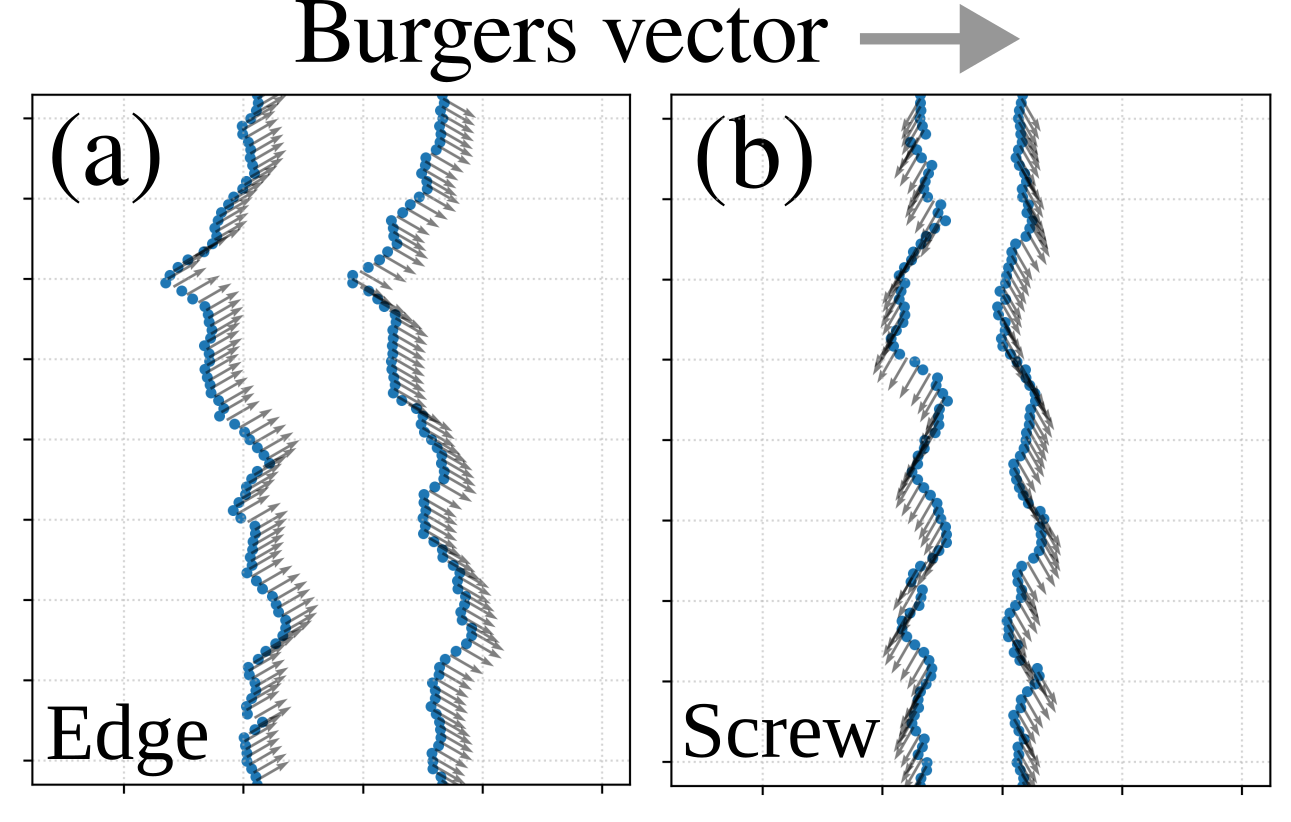}
    \caption{Shape of dislocations under correlated SF noise. Burgers vectors of each segment are visualized with the grey arrows.}
    \label{fig:disloc_shape_sf}
\end{figure}

Figure \ref{fig:disloc_shape_sf} presents the relaxed configurations of dislocation lines under correlated SF noise for each character type. As predicted by the classical isotropic model \cite{hirsch1965stacking}, the edge character exhibits a larger separation distance between partial dislocations compared to the screw dislocation. We observe that the shape of the two partial dislocations in both cases is strongly coupled. This suggests the existence of a critical separation length—as previously conceptualized by Vaid et al. \cite{vaid2022pinning}—at which the interaction force overcomes the pinning contribution of the noise potential.


Next, we attempt to justify the character dependence of the CRSS for both correlated and uncorrelated noises: the edge character exhibits higher CRSS than the screw character. Our preliminary tests on the effect of line tension showed that a stronger line tension (screw character) consistently decreases the CRSS. This is because higher line tension helps the dislocation segment overcome the noise potential barriers. The lower core energy of the screw dislocation (Figure \ref{fig:char_depend}(d)) suggests that it resists the perturbation more strongly than the edge dislocation, which is reflected in its lower CRSS (Figure \ref{fig:char_depend}(d)).

\begin{figure}[H]
    \centering
    \adjincludegraphics[,width=0.7\textwidth,trim={{.0\width} {.0\width} {.0\width} {.0\width}},clip]{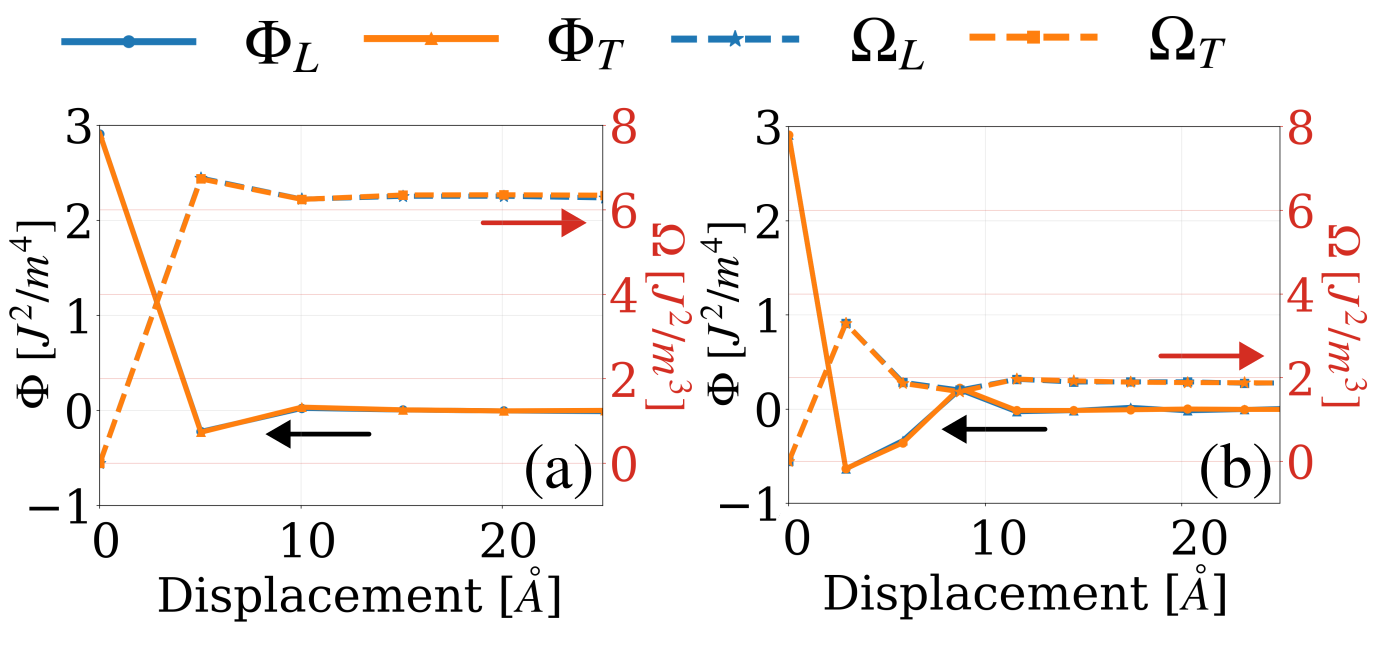}
    \caption{ISF correlation ($\Phi$) and its integral along the line-tangent direction ($\Omega$) for (a) edge character and (b) screw character, with the black and red arrows indicating the $y$-axis for the correlation and its integral, respectively.}
    \label{fig:corint_tangent}
\end{figure}

Reference \cite{rida2022influence} showed that, for screw dislocations in FCC, the integral of the correlation is directly proportional to the CRSS. Here, we attempt to apply this same reasoning. The raw correlation values along the dislocation line tangent for each configuration (shown in Figures \ref{fig:char_depend}(a)\&(b), and in Figures \ref{fig:corint_tangent}(a)\&(b)) are presented as $\Phi$, and their integrals as $\Omega$ in Figure \ref{fig:corint_tangent}. We observe that the correlations exert a stronger influence on edge character than on screw character, reflected in the larger CRSS drop for the edge case than for the screw case (Figure \ref{fig:char_depend}(c)).


This behavior can be attributed to the combined effects of character-dependent line tension and numerical effects: the interaction energy assigns a lower line tension to edge character, and the grid spacing $\Delta x$ for edge dislocations is larger than that for screws because its tangent lies along a $\langle 112 \rangle$ direction, which for the uncorrelated case was shown to increase the CRSS \cite{rida2022influence}. These combined effects amplify the strengthening for the uncorrelated edge case.



%% file: sections/discussion.tex
\subsection{Origin of the SFE correlation}\label{sec:DiscussSub1}

In this section, we attempt to investigate how the negative correlation in the dilute limit originates from displacement effects.


\begin{figure}[H]
    \centering
    \includegraphics[width=0.9\linewidth]{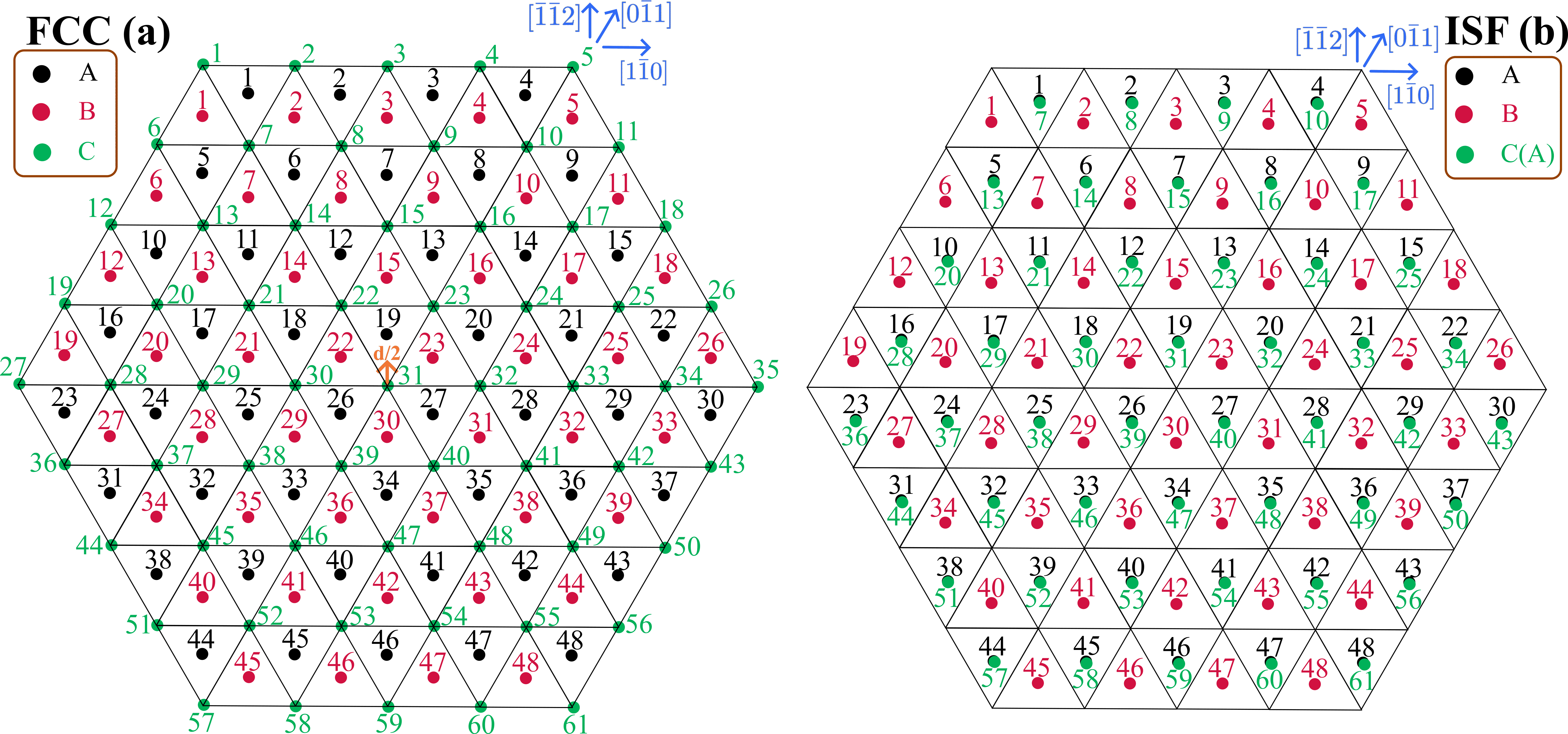}
    \caption{Schematic comparison between perfect FCC structure (left) and faulted structure (right) following C-layer displacement along the $[\bar{1}\bar{1}2]$ direction. The red arrow indicates the displacement direction. The C layer, colored green, depicts A stacking after the displacement but is labeled as C for mathematical clarity.}
    \label{fig:three_layer_drawing}
\end{figure}


The conventional FCC stacking sequence follows a repeating ABC pattern, as illustrated in the left panel of Figure \ref{fig:three_layer_drawing}. For model clarity, the schematic's out of plane direction is aligned with the $[111]$ crystallographic direction. 
The energy of a given lattice site is approximated as a sum of bond energies over the neighboring sites:
\begin{equation}
    E_{i} = \frac{1}{2}\left [ \sum _{\{j\}}e_{j}^{X(n)}\right ]
\end{equation}
\noindent where $e_{\{j\}}^{X(n)}$ denotes bond energies with the superscript $X$ representing the layer of the end site of the bond assuming site $i$ as the origin and the number in parentheses denoting nearest neighbor (NN) order, and $\{j\}$ is the set of lattice sites involved in the bonding. The neighbor orders used in this section are defined as follows:

\begin{equation}
    n =
    \begin{cases}
        1 & \text{1NN in FCC}\\
        2 & \text{2NN in FCC}\\
        3 & \text{3NN in HCP}\\
        4 & \text{3NN in FCC} 
    \end{cases}
\end{equation}

This convention resolves ambiguity in faulted structures by distinguishing third-neighbor shells in the faulted configuration from those in the perfect FCC configuration, as illustrated in Figure \ref{fig:three_layer_drawing}.




Using lattice point 31 in layer C as the reference point, we define its energy as the weighted sum of bond energies between atom pairs with a cutoff of $a_0\sqrt{\frac{3}{2}}$, i.e., third nearest neighbors in the FCC crystal structure. We will compute increments of energy, and hence, we will only consider bonds that might be broken in the displacement, i.e., a broken bond algorithm. Therefore, the energy at site 31 before the displacement is calculated as 


\begin{equation}
    \bar{E}_{31} = \frac{1}{2}\left [ \sum _{\{i\}}e_{i}^{B(1)}+\sum _{\{j\}}e_{j}^{B(2)}+\sum _{\{k\}}e_{k}^{A(4)}+\sum _{\{l\}}e_{l}^{B(4)}\right ]
\end{equation}



Around site 31 we have $\{i\}=\{22,23,30\}$, $\{j\}=\{15,29,31\}$, $\{k\}=\{19,26,27\}$, and $\{l\}=\{14,16,21,24,36,37\}$. In a monoatomic system, all bond energies are equal for a given distance. However, for a random solid solution, the energy of each bond fluctuates stochastically based on the atomic species forming the bond. For a random solution the bond energies are uncorrelated and identically distributed with mean $\mu_i$ and variance $V_i$. The mean and variance are defined as

\begin{equation}
    \begin{aligned}
        &cov(e_i, e_j) = 0 \implies \EX\left[e_i e_j\right] = \EX\left[e_i\right] \EX\left[e_j\right] \\
        &\mu_k = \EX \left [ e_i^{X(k)}\right ] \\
        &V_k = \EX \left [\left (e_i^{X(k)}\right )^2 \right ] - \left ( \EX \left [ e_i^{X(k)}\right ] \right )^2
    \end{aligned}
\end{equation}

The SFE is related to the expected value of the difference in energy at a lattice site after the C layer is displaced in the $[\bar{1}\bar{1}2]$ direction (see Fig~\ref{fig:three_layer_drawing}). Note that this displacement changes the stacking sequence from ABC to ABAB. We can express the energy of the faulted configuration following the same bonding scheme. Therefore, the energy at the lattice point 31 after the displacement is given by:



\begin{equation}
    E_{31} = \frac{1}{2}\left [ \sum _{\{i\}}e_{i}^{B(1)}+\sum _{\{j\}}e_{j}^{B(2)}+\sum _{\{k\}}e_{k}^{A(3)}+\sum _{\{l\}}e_{l}^{B(4)}\right ]
\end{equation}



With the corresponding lattice sites $\{i\} = \{15,22,23\}$, $\{j\} = \{14,16,30\}$, $\{k\} = \{19\}$, and  $\{l\} = \{8,9,21,24,29,31\}$. 


The change in energy at site 31 on layer C is therefore:

\begin{equation}
    \begin{aligned}
   \Delta E_{31} &= E_{31} - \bar{E}_{31} \\
   &= \frac{1}{2}\left [ \Delta e_i^{B(1)} + \Delta e_j^{B(2)}+\Delta e_k^{A(3)}+\Delta e_l^{A(4)}+\Delta e_{m}^{B(4)} \right ]
    \end{aligned}
    \label{eq:DeltaE}
\end{equation}


where each term is defined as follows:

\begin{equation}
\begin{aligned}
    \Delta e_i^{B(1)} &= e_{15}^{B(1)}-e_{30}^{B(1)} \\
    \Delta e_j^{B(2)} &= e_{14}^{B(2)}+e_{16}^{B(2)}+e_{30}^{B(2)}-e_{15}^{B(2)}-e_{29}^{B(2)}-e_{31}^{B(2)} \\
    \Delta e_k^{A(3)} &= e_{19}^{A(3)} \\
    \Delta e_l^{A(4)} &= -e_{19}^{A(4)}-e_{26}^{A(4)}-e_{27}^{A(4)} \\
    \Delta e_m^{B(4)} &= e_{8}^{B(4)}+e_{9}^{B(4)}+e_{29}^{B(4)}+e_{31}^{B(4)}-e_{14}^{B(4)}-e_{16}^{B(4)}-e_{36}^{B(4)}-e_{37}^{B(4)}
    \label{eq:Comp}
\end{aligned}
\end{equation}

Using the same approach, we can calculate $\Delta E_{47}$, as a point in the direction of the displacement, $\Delta E_{30}$ for a point in the $\left [ 1 \bar{1} 0\right ]$ direction perpendicular to the displacement, $\Delta E_{38}$ for a point in a different $\langle 112 \rangle$ direction, and $\Delta E_{39}$ for a point in a different $\langle 110 \rangle$ direction. Following a similar methodology (Eqs.~\ref{eq:DeltaE} and \ref{eq:Comp}), we can compute the values for $\Delta E_{47}$, $\Delta E_{30}$, $\Delta E_{38}$ and $\Delta E_{39}$. The derivation details are provided in the Supplementary Material.

From these expressions we obtain:
\begin{gather}
    \EX(\Delta E_i) = \frac{1}{2}\left ( \mu _3 - 3\mu_4 \right )
\end{gather}

The expected value of $\Delta E_i$ corresponds to half the SFE, since there will be another term like this for the bonds broken and reformed from planes B and A above C. Hence, we can write, $\gamma_{ISF} = \mu_3 - 3\mu_4$. Note that the average bond energies $\mu$ are negative, and the distance between atoms in $(3)$ and $(4)$ bonds (third nearest neighbor in ideal HCP, $2a_0/\sqrt{3}$, and third nearest neighbors in FCC, $a_0\sqrt{\frac{3}{2}}$) is small, and hence we do not expect a large difference in their absolute values, which will result in a positive $\gamma_{ISF}$.

To understand the origin of the correlations that we obtain from atomistic simulations, we compute the covariance between the SFEs of different sites. The covariance along the $\left [ \bar{1} \bar{1} 2 \right ]$ direction (the displacement direction) is given by

\begin{align}
    cov(\Delta E_{31}, \Delta E_{47}) &= \EX \left[ \Delta E_{31}\Delta E_{47}\right] -  \EX\left[\Delta E_{31}\right]\EX\left[\Delta E_{47}\right] \\ \nonumber
    &= \frac{1}{4}\left [ -V_1 -3V_2 -2V_4  \right ]
\end{align}
which is always negative, as it was observed in the atomistic results. In the $\left [ 1 \bar{1} 0 \right ]$ direction, perpendicular to the displacement, we obtain
\begin{gather}
    cov(\Delta E_{31}, \Delta E_{30}) = \frac{1}{4}\left [ -4V_2+3V_4 \right ]
\end{gather}

In this case, the value can be positive or negative depending on the variance of the bond energies between the 2NN and 3NN in the FCC lattice. In the atomistic results, the values close to the origin are positive, indicating that the variance of the latter ($V_4$) is slightly larger than the first ($V_2$).

In a different $\langle 112 \rangle$ direction, we find
\begin{gather}
    cov(\Delta E_{31}, \Delta E_{38}) = \frac{1}{2}\left [ -V_4 \right ]
\end{gather}
which is negative as observed from the results with the interatomic potential.

And finally, in a different $\langle 110 \rangle$ direction, we obtain
\begin{gather}
    cov(\Delta E_{31}, \Delta E_{39}) = \frac{1}{4}\left [ -V_4 \right ]
\end{gather}
which would be also negative. The atomistic results show more complex trends at distances larger than those captured by the pair interaction approach, although with negative values for short distances, in agreement with the broken bond model.

\begin{figure}[H]
    \centering
    \includegraphics[width=0.8\linewidth]{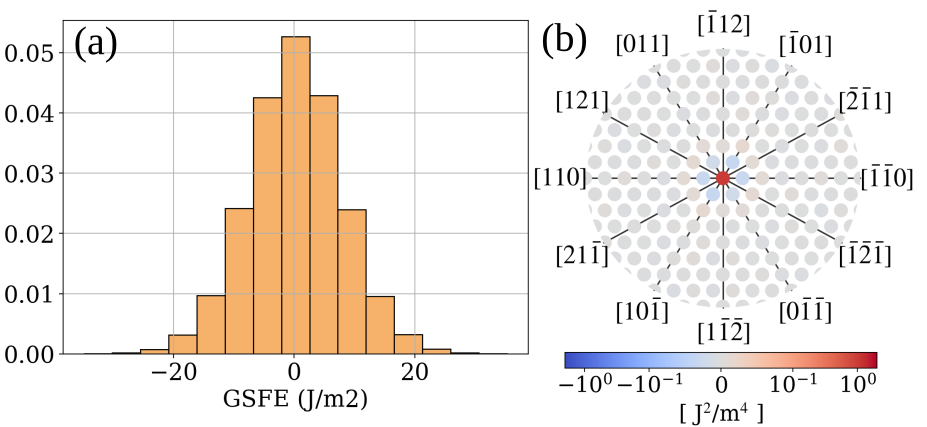}
    \caption{(a) GSFE distribution of the resampled random solid solution and (b) its spatial correlation (b). The resampled correlation trend suggests that the energy distribution exhibits characteristics of Gaussian white noise.}
    \label{fig:resampled_correlation}
\end{figure}



The SFE broken-bond model discussed here resembles the SS model described in Ref. \cite{nag2024solute}. Our model provides important insights into the source of the correlations. In particular, it captures the negative correlation along the displacement direction and the main trends along other directions. We also tested the effect of the initial atomic configuration on the correlations. Instead of keeping the solute atoms in their corresponding positions after displacement, we resampled their positions randomly and recomputed the $\gamma_{ISF}$ at each grid point. The results are shown in Fig.~\ref{fig:resampled_correlation}, where the correlation values exhibit characteristics of Gaussian white noise. This provides evidence for the hypothesis that the correlations arise from displacing the solute atoms when the leading partial dislocation glides to create a stacking fault.

\subsection{Interplay between SS and SFE fluctuations}\label{sec:DiscussSub2} 
\begin{table}[htbp]
\centering
\begin{tabular}{lllll}
\hline
Type & Symbol & Value & Unit & Ref. \\
\hline\hline
SS Noise Variance & $\langle\tau^2_{p}\rangle_{Mg5}$ & 0.16 & $\text{GPa}^2$ & \cite{geslin2021microelasticity, geslin2021microelasticity2} \\
 & $\langle\tau^2_{p}\rangle_{Mg10}$ & 0.32 & $\text{GPa}^2$ & \cite{geslin2021microelasticity, geslin2021microelasticity2} \\
 & $\langle\tau^2_{p}\rangle_{Mg15}$ & 0.43 & $\text{GPa}^2$ & \cite{geslin2021microelasticity, geslin2021microelasticity2} \\
\hline\hline
\end{tabular}
\caption{Noise variances are computed based on the analytical correlation expressions provided by Geslin and Rodney~\cite{geslin2021microelasticity, geslin2021microelasticity2}.}
\label{Table:dddparams_combined}
\end{table}

The combined effect of ISF and SS noises on the CRSS is investigated in this section. The initial assumption is that the noise terms are independent and superposition holds, i.e. the cross correlations between ISF and SS fluctuations are negligible. Deeper analysis on such correlations is left for a future work. The analytical SS correlation expressions for AlMg$_x$ alloys are given in Refs. \cite{geslin2021microelasticity, geslin2021microelasticity2}, and the same random grating scheme is utilized to sample the SS noise field in DDD as described in Section \ref{sec:MethodSub3}. The sampled SS noise field is superimposed onto the existing ISF noise to add the pinning stress generated from solute misfit energy.

\begin{figure}[H]
    \centering
    \adjincludegraphics[,width=0.7\textwidth,trim={{.0\width} {.0\width} {.0\width} {.0\width}},clip]{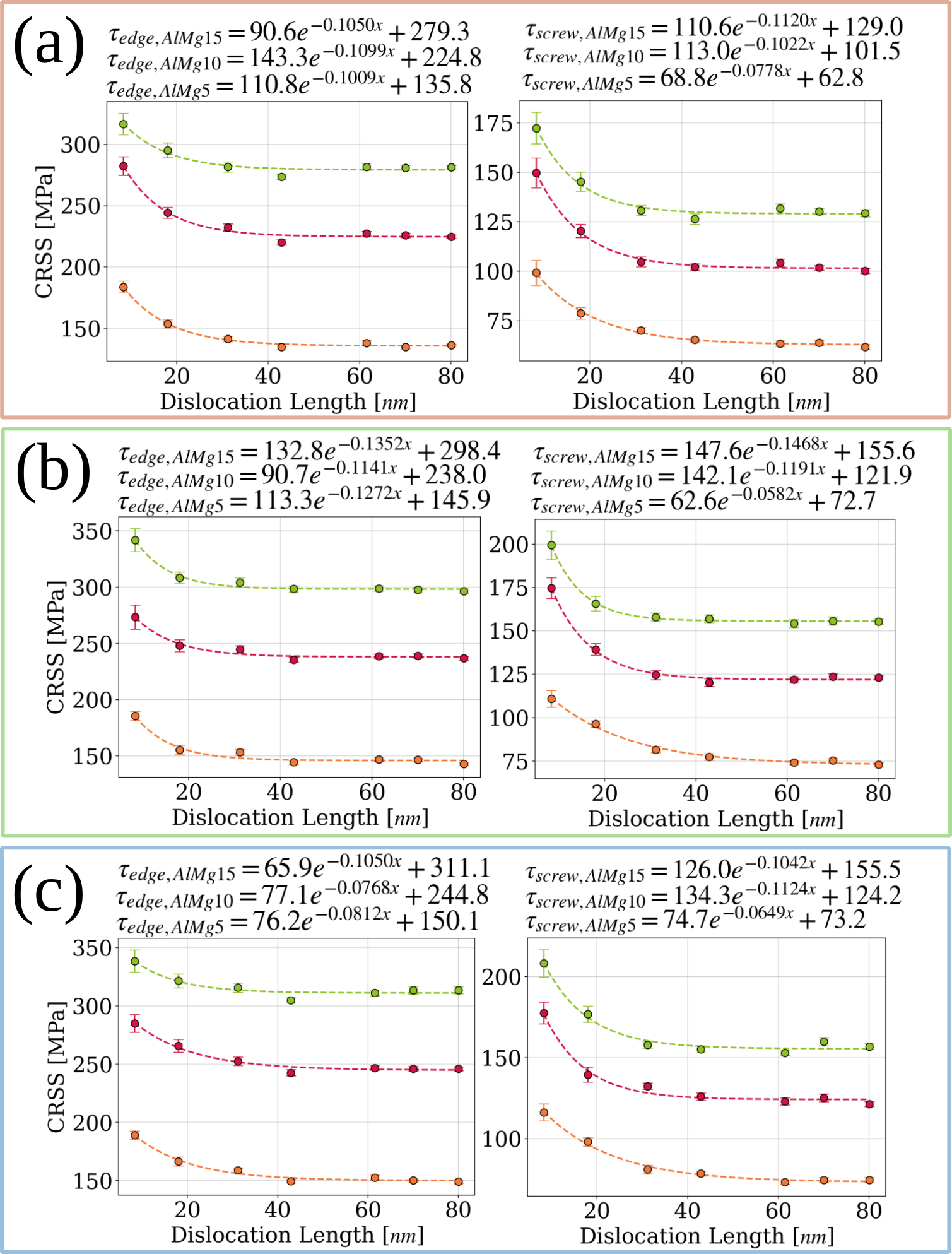}
    \caption{CRSS convergence as a function of dislocation lengths, with edge dislocations in the left column and screw dislocations in the right column. (a) Correlated SS and constant SFE, (b) correlated SS and correlated SFE noise, and (c) correlated SS and uncorrelated SFE noise.}
    \label{fig: combconv}
\end{figure}

The analysis examines convergence in the CRSS as a function of dislocation length using Section \ref{sec:MethodSub3}'s detection scheme and Section \ref{sec:MethodSub4}'s glide control parameters, testing all partial dislocation noise combinations: (a) correlated SS noise with constant SFE, (b) correlated SS fluctuations and correlated SF noise, and (c) correlated SS noise with uncorrelated SF noise (Figure \ref{fig: combconv}). Controlling glide distance reveals consistent CRSS decay trends as a function of dislocation length, matching those in Section \ref{sec:ResultSub2}. Exponential CRSS decay models extrapolate to infinite-length dislocation behavior across noise combinations. CRSS values for case (a) are lower than those for cases (b) and (c) due to the strengthening effect induced by the local SFE fluctuations. 


In general, the edge character exhibits significantly higher CRSS values than the screw character, indicating that edge components interact more strongly with SS noise. It is known that screw character interacts with dilatational solutes in a lesser extent compared to edge character. Specifically, the interaction becomes zero at infinite length due to the property $\int \Sigma_L dx = 0$~\cite{rida2022influence}. This proves that the correlated noise model is applicable not only to simple dissipative equations but also to dislocation lines modeled as physical loop structures, consistent with the findings of Ref. \cite{rida2022influence}.

\begin{table}[htbp]
\centering
\begin{tabular}{llll}
\toprule
Mg at-\%  & Character & Correlated [MPa] & Uncorrelated [MPa] \\
\midrule
5  & Edge    & 10.1 & 14.3 \\
10 & Edge    & 13.2 & 20.0 \\
15 & Edge    & 19.1 & 31.8 \\
5  & Screw   & 9.9  & 10.4 \\
10 & Screw   & 20.4  & 22.7 \\
15 & Screw   & 26.6  & 26.5 \\
\bottomrule
\end{tabular}
\caption{CRSS deviation magnitudes for cases (b) and (c) relative to case (a), shown as functions of Mg concentration and dislocation character.}
\label{tab:crss_deviation}
\end{table}



We further quantified the correlation effect on the SF strengthening contribution by computing the difference between the CRSS values fitted at the infinite-length limit for case (a) and those for cases (b) and (c). This difference—defined as the CRSS offset from the cases with SF noise to the case without SF noise—increases as the Mg concentration rises from 5\% to 15\% (Table \ref{tab:crss_deviation}). As shown in the table, case (c) (uncorrelated) generally exhibits larger CRSS increases than case (b) (correlated), and the screw character shows closer CRSS values between the correlated and uncorrelated noise compared to the edge character. However, it is worth noting that the difference in the CRSS with correlated ISF noise is slightly larger for screw than edge. This might be due to numerical reasons (lack of statistics, precision) or to some interplay between noise sources that will be investigated in a subsequent work.




\begin{figure}[H]
    \centering
    \adjincludegraphics[,width=0.8\textwidth,trim={{.0\width} {.0\width} {.0\width} {.0\width}},clip]{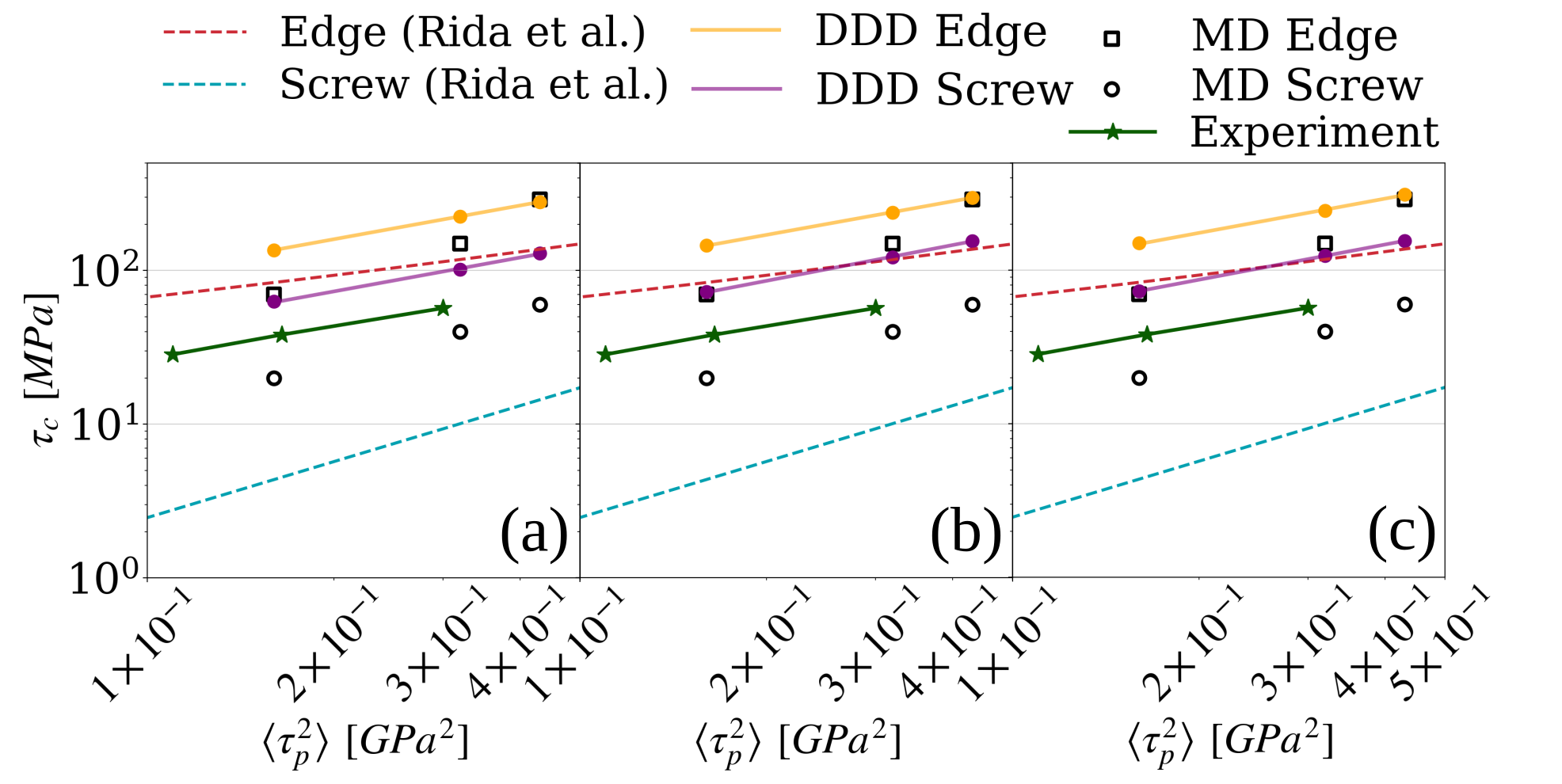}
    \caption{Comparison of CRSS data fitted to infinitely long dislocations with the model by Rida et al.~\cite{rida2022influence} (dashed lines). (a) SS with constant SF noise, (b) SS with correlated SF noise (c) SS with uncorrelated SF noise.}
    \label{fig: infind2}
\end{figure}

CRSS values extrapolated to infinite dislocation lengths for cases (a), (b), and (c) were then compared against the force-based model from Ref.~\cite{rida2022influence} in Figure~\ref{fig: infind2}. Combined SS/SF noise simulations overestimate CRSS relative to both edge- and screw-character force-based models. While some discrepancy was expected due to this work's focus on partial dislocations versus force-based models' assumption of perturbations on full dislocation lines, the observed deviation magnitude exceeded initial predictions. This discrepancy arises from fundamental differences in nodal-based DDD implementations: our MoDELib-based simulations employ three-dimensional parametric curves discretized with cubic Hermite splines~\cite{po2014recent}, which interpolate both position and derivative information at segment endpoints. This approach computes line tension forces in full 3D space, where stabilizing forces contain $x$, $y$, and $z$ components.
In contrast, Ref. \cite{rida2022influence} uses direct time evolution of individual nodes with spline-interpolated noise, assigning line tension contributions solely along the glide direction and constant along the line for screw and edge, i.e., independent of the curvature of the line. These methodological differences—particularly the treatment of dislocation line tension in three dimensions versus single-direction approximation and the explicit treatment of partial dislocations—lead to the variations observed in the results. 
We also observe that DDD with both sources of noise moderately overestimates the MD values. The reason for this discrepancy is probably three-fold: inaccuracies in the calculation of the CRSS, correlations in the SS and SF noises neglected in this study, and core effects. These discrepancies will be the topic of a future work.

%% file: sections/conclusion.tex
This work provides a tool to systematically investigate the individual contributions of local SFE fluctuations and SS noise to the strengthening effect in FCC crystals. We focus on quantifying the strengthening mechanisms in crystalline solids through locally fluctuating SF energetics and their superposition onto the SS strengthening effect. Spatial autocorrelations of SFE, extracted from MS simulations, are successfully integrated into continuum DDD frameworks to isolate individual contributions to dislocation behavior. We observe that the fluctuations in the SFE always increase the CRSS, for both screw and edge dislocation characters. Compared to SS noise, the increase in strength is moderate. Furthermore, we have shown that the fluctuations in the SFE are spatially correlated, with negative correlations along the slip direction and more complex relations in other directions. A pair interaction model is able to recover the origin of the negative correlations and the fact that in other directions the sign depends on the strength of the interactions. These negative correlations decrease the CRSS compared to uncorrelated fluctuations.  Interestingly, although SFE fluctuations do not depend on the dislocation character, we observe edge dislocations with slightly larger CRSS than screw. The reason seems to lie on the fact that the line tension for screws is higher, which helps overcoming energetic barriers, and with the fact that the grid spacing (following an FCC structure on a $(111)$ plane) is larger in the edge orientation. The latter is a numerical artifact, while the first is physically based.

While this framework enables systematic analysis of local SF fluctuations, three limitations require attention:

\begin{itemize}
    \item DDD scaling parameters (line tension coefficients and non-singular regularization lengths) were selected to circumvent ultraviolet limitations~\cite{boleininger2020ultraviolet} rather than derived from atomistic simulations. Future work should establish these parameters through MD-calibrated approaches to strengthen the multi-scale framework.
    \item The presence of a SF modifies the atomic configuration potentially varying the SS stress correlations~\cite{nohring2017dislocation}, yet our model assumes linear superposition of SS and SF noise effects—both increasing dislocation strength. The validity of treating these noise sources as wholly independent requires rigorous verification.
    \item Energetic differences between leading and trailing dislocations in the cases of mixed characters induced by SF noise necessitate expanded investigation across broader alloy compositions.
\end{itemize}

Resolving these challenges—particularly the non-trivial decoupling of noise effects—will be addressed in subsequent work to establish robust superposition principles for multi-noise dislocation dynamics.

%% file: sections/appendix.tex
\appendix

\section{Bisection Searching}\label{sec:app_bisection}


This procedure, termed "bisection searching" for its algorithmic similarity, follows these steps:

\begin{enumerate}
    \item Apply an initial stress estimate likely to initiate dislocation glide.  
    \item If the time-averaged plastic strain rate $\langle|\dot{\beta}^p|\rangle = 0$ (computed over the final three snapshots), the stress is below CRSS.  
    \item If $\langle|\dot{\beta}^p|\rangle > 10^{-10}$ $c_s^{-1}$, the stress exceeds CRSS.  
    \item Adjust stress iteratively:
    \begin{itemize}
        \item Increase stress by a fixed increment when below CRSS  
        \item Halve the stress interval when above CRSS 
    \end{itemize}
    \item Terminate when $|\tau^+_{CRSS} - \tau^-_{CRSS}| < 5$–$6$ MPa, where $\tau^+_{CRSS}$ is defined by: 
    \begin{equation}
    \begin{aligned}
&\tau^+_{CRSS} = \tau_{app} \\ \text{if} \quad &\langle|\dot{\beta}^p|\rangle > 10^{-10} [c_s^{-1}] \quad \text{and} \quad \tau^+_{CRSS} - \tau^-_{CRSS} < 5\text{-}6  \text{MPa}
\label{eq:crss}
    \end{aligned}
    \end{equation}
 \end{enumerate}

\section{Calculation of Correlated Random Noise}\label{sec:app_correlated_noise}
The sampled fluctuating field, $\psi({\bm{x}_{(m,n)}})$, is assumed to be a stationary, Gaussian stochastic process, meaning it exhibits the following statistical properties:

\begin{equation}
    \langle \psi(\bm{x}_{(m,n)}) \rangle = 0
    \label{eq:}
\end{equation}

\begin{equation}
    \langle \psi(\bm{x}_{(m,n)}) \psi(\bm{x}_{(m+p,n+q)}) \rangle = \eta^2 \sum^{+\infty}_{i=-\infty} \sum^{+\infty}_{-\infty} W_{(m,n)} W_{(m-p,n-q)}
    \label{eq:}
\end{equation}

where $W_{(m,n)}$ is the weight factor at grid site $(m,n)$, and $\eta^2$ is the mean-square deviation from the zero mean. According to the random grating method, the stationary Gaussian process $\psi_{(m,n)}$ can be expressed as:

\begin{equation}
    \psi({\bm{x}_{(m,n)}}) = \sum^{+\infty}_{m=-\infty} \sum^{+\infty}_{n=-\infty} W_{(m,n)}X_{(m+p,n+q)}
    \label{eq:}
\end{equation}

where $X_{(m,n)}$ for $m,n \in \mathbb{Z}$ is a sequence of independent, Gaussian random variables with zero mean and unit variance:

\begin{align}
    \langle X_{(m,n)} \rangle &= 0 \\
    \langle X_{(m,n)} X_{(p,q)} \rangle &= \eta_{mp} \eta_{nq}
\end{align}

The weight factors $W_{(m,n)}$ are determined from the target autocorrelation function $R(\mathbf{x})$ via the Wiener–Khinchin theorem, which relates the autocorrelation to its power spectral density $\widehat{R}(|\bm{k}|)$:

\begin{equation}
    \sum^{+\infty}_{m=-\infty} \sum^{+\infty}_{n=-\infty} W_{(m,n)} W_{(m-p,n-q)} = \int \widehat{R}(|\bm{k}|)e^{i\bm{k}\cdot\bm{x}_{(n,m)}}\frac{d^2\bm{k}}{(2\pi)^2}
    \label{eq:}
\end{equation}
where $\bm{k} = (k_x, k_y, k_z)$ and $k_x = 2\pi n/L_x$ and $k_y = 2\pi m /L_y$ are the wave vector components with $n$ and $m$ integers. The $\widehat{R}(|\bm{k}|)$ is the power spectral density of the autocorrelation function computed in this work:

\begin{equation}
    \widehat{R}(|\bm{k}|) = \int R(|\bm{x}|)e^{-i\bm{k}\cdot\bm{x}} d^2\bm{x}
    \label{eq:}
\end{equation}

We introduce the representation of the weight factor in Fourier space.
\begin{equation}
    W_{(m,n)} = \int \widehat{W} e^{i\bm{k}\cdot{\bm{x}}}\frac{d^2\bm{k}}{(2\pi)^2}
    \label{eq:}
\end{equation}

In the continous limit $\Delta x \to 0$ and $\Delta y \to 0$, the weight factor related to:
\begin{equation}
    \frac{1}{\Delta x} \frac{1}{\Delta y} \int \widehat{W}(\bm{k})\widehat{W}(-\bm{k})e^{i\bm{k}\cdot\bm{x}_{(n,m)}}\frac{d^2\bm{k}}{(2\pi)^2} = \int R(|\bm{k}|)e^{i\bm{k}\cdot\bm{x}_{(n,m)}} \frac{d^2\bm{k}}{(2\pi)^2}
    \label{eq:}
\end{equation}

The continous assumption justifies the fact that the $\widehat{W}$ is real. Therefore, the $\widehat{W}$ is described by the following relation
\begin{equation}
    \widehat{W}(\bm{k}) = \sqrt{\Delta x \Delta y} \sqrt{R(|\bm{k}|)}
    \label{eq:}
\end{equation}

We acquire the expression for the weight factor by taking inverse Fourier transform.
\begin{equation}
    W_{(m,n)} = \sqrt{\Delta x \Delta y} \int \sqrt{R(|\bm{k}|)} e^{i\bm{k} \cdot \bm{x}_{(m,n)}} \frac{d\bm{k}^2}{(2\pi)^2}
    \label{eq:weight_fourier}
\end{equation}

The random field $\psi_{(m,n)}$ defined on an $M \times N$ grid is assumed to be periodic. Under this assumption, $\psi_{(m,n)}$ can be expressed in terms of the weight factors $W_{(m,n)}$ and independent Gaussian random variables $X_{(m,n)}$, both of which inherit the same periodicity:

\begin{equation}
    W_{(m,n)} = \frac{1}{\sqrt{MN}} \sum^{M/2-1}_{p=-M/2} \sum^{N/2-1}_{q=-N/2} \widehat{W}_{(m,n)} e^{i(2\pi mp/M+ 2\pi nq/N)}
    \label{eq:w_ft_rel}
\end{equation}

\begin{equation}
    X_{(m,n)} = \frac{\eta}{\sqrt{MN}} \sum^{M/2-1}_{p=-M/2} \sum^{N/2-1}_{q=-N/2} \widehat{X}_{(m,n)} e^{i(2\pi mp/M+ 2\pi nq/N)}    \label{eq:x_ft_rel}
\end{equation}

The inverse discrete Fourier transform of Eq.~\eqref{eq:w_ft_rel} is given by:

\begin{equation}
    \widehat{W}_{(m,n)} = \frac{1}{\sqrt{MN}} \sum^{M/2-1}_{-M/2} \sum^{N/2-1}_{-N/2} W_{(p,q)} e^{-i(2\pi mp/M + 2\pi nq/N)}
    \label{eq:}
\end{equation}

In the continuous limit ($\Delta x, \Delta y \to 0$), this becomes:
\begin{equation}
    \widehat{W}_{(m,n)} = \frac{1}{\sqrt{MN}\Delta x \Delta y} \int^{\infty}_{-\infty} W{(\bm{x})} e^{-i(\frac{2\pi m x}{\mathcal{L_x}} + \frac{2\pi ny}{\mathcal{L_y}})} d^2\bm{x}
    \label{eq:}
\end{equation}

where $\mathcal{L}_x = M \Delta x$ and $\mathcal{L}_y = N \Delta y$ are the physical dimensions of the simulation cell in the $x$ and $y$ directions, respectively. Using the Wiener–Khinchin relation, the Fourier-space weight $\widehat{W}_{(p,q)}$ can be directly related to the power spectral density $\widehat{R}(|\bm{k}|)$. This leads to the following expression:

\begin{equation}
    \widehat{W}_{(m,n)} = \frac{1}{\sqrt{M N\Delta x \Delta y}} \sqrt{C(|\bm{k}|)}
    \label{eq:final_weight_fourier}
\end{equation}

We define the Gaussian random variable as:

\begin{equation}
    \widehat{X}_{(m,n)} = \frac{1}{\sqrt{2}}(\widehat{\mathcal{Q}}_{(m,n)} + i\widehat{\mathcal{P}}_{(m,n)})
    \label{eq:final_gauss_rel}
\end{equation}

where $\widehat{\mathcal{Q}}_{(p,q)}$ and $\widehat{\mathcal{P}}_{(p,q)}$ are independent, real-valued Gaussian random variables with zero mean and unit standard deviation in Fourier space. Combining Eqs.~\eqref{eq:final_weight_fourier} and~\eqref{eq:final_gauss_rel}, the correlated random field $\psi(\bm{x}_{(m,n)})$ is constructed as:

\begin{equation}
\begin{aligned}
    &\psi(\bm{x}_{(m,n)}) = \\
    &\frac{\eta}{\sqrt{MN\Delta x \Delta y}} \sum^{M/2-1}_{q=-M/2} \sum^{N/2-1}_{p=-N/2} \frac{1}{\sqrt{2}}(\widehat{\mathcal{Q}}_{(m,n)} + i\widehat{\mathcal{P}}_{(m,n)}) \sqrt{\widehat{R}(|\bm{k}|)}e^{i\bm{k}_{(q,p)} \cdot \bm{x}_{(q,p)}}
    \label{eq:}
\end{aligned}
\end{equation}

\section{Standard deviation Correction}\label{sec:std_pad_correction}

\begin{figure}[H]
    \centering
    \adjincludegraphics[,width=1\textwidth,trim={{.0\width} {.0\width} {.0\width} {.0\width}},clip]{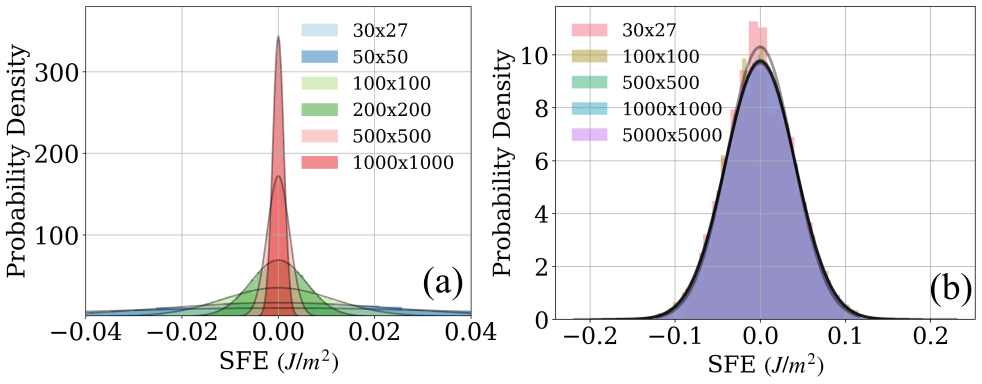}
    \caption{(a) Sampled SF noise distribution for each grid size with zero padding (before energy normalization); (b) Sampled SF noise distribution after energy normalization}
    \label{fig:zero_pad_correction}
\end{figure}

For correlation calculations, spatial correlation fields are zero-padded to increase the sampled noise domain size. However, modifying the discrete data size introduces artifacts due to the nature of discrete signals. Specifically, the total energy (sum of squared values) of the original spatial correlation field must be preserved after zero-padding around the center. In Figure~\cite{fig:zero_pad_correction}, the $30 \times 27$ grid size shown in panel (a) represents the unpadded random noise sample. As zero padding is introduced and padding size increases, the standard deviation decreases and the distribution narrows due to spectral interpolation effects. Consequently, spatial correlation energy is normalized when zero padding is applied to accommodate larger glide plane sizes in DDD simulation configurations.

This energy conservation requirement stems from the fact that zero-padding performs sinc interpolation in the frequency domain without adding new information. Let the energy of random function $f_r(x)$ be $E$, with $N_0$ and $N$ representing the discrete data sizes before and after padding, respectively. The standard deviation after zero padding relates to the original through the grid size ratio:

\begin{equation}
    \sigma = \sigma_0 \sqrt{\frac{N_0}{N}}
    \label{eq:}
\end{equation}
where $\sigma$ is the post-padding standard deviation and $\sigma_0$ is the original value. This relationship arises because zero-padding interpolates the spectrum without changing the underlying signal energy, effectively distributing the same energy across more samples. Therefore, the correction factor $\sqrt{N_0/N}$ must be applied to the noise field whenever zero padding is introduced to maintain energy conservation in spatial correlation data.

